# Classification of NEQR Processed Classical Images using Quantum Neural Networks (QNN)

**Santanu Ganguly***

* Cisco Systems International, UK

ORCID iD: https://orcid.org/0000-0003-0141-9228

***Abstract-*** A quantum neural network (QNN) is interpreted today as any quantum circuit with trainable continuous parameters. This work builds on previous works by the authors and addresses QNN for image classification with Novel Enhanced Quantum Representation of (NEQR) processed classical data where Principal component analysis (PCA) and Projected Quantum Kernel features (PQK) were investigated previously by the authors as a path to quantum advantage for the same classical dataset. For each of these cases the Fashion-MNIST dataset was downscaled using PCA to convert into quantum data where the classical NN easily outperformed the QNN. However, we demonstrated quantum advantage by using PQK where quantum models achieved more than ~90% accuracy surpassing their classical counterpart on the same training dataset as in the first case. In this current work, we use the same dataset fed into a QNN and compare that with performance of a classical NN model. We built an NEQR model circuit to pre-process the same data and feed the images into the QNN. Our results showed marginal improvements (only about ~5.0%) where the QNN performance with NEQR exceeded the performance of QNN without NEQR. We conclude that given the computational cost and the massive circuit depth associated with running NEQR, the advantage offered by this specific Quantum Image Processing (QIMP) algorithm is questionable at least for classical image dataset. No actual quantum computing hardware platform exists today that can support the circuit depth needed to run NEQR even for the reduced image sizes of our toy classical dataset.

## 1 Introduction

Quantum machine learning (QML) is a cross-disciplinary subject made up of two of the most exciting research areas: quantum computing and classical machine learning. Classical deep neural networks are efficient tools for machine learning and serve as the fundamental base for the development of deep quantum learning methods. A quantum neural network (QNN) is interpreted today as any quantum circuit with trainable continuous parameters. Quantum computation (QC) and quantum information are disciplines that have appeared in various areas of computer science, such as information theory [1], cryptography [24], emotion representation and image processing [25], because tasks that appear inefficient on classical computers can be achieved by exploiting the power of QC [1, 2-5].

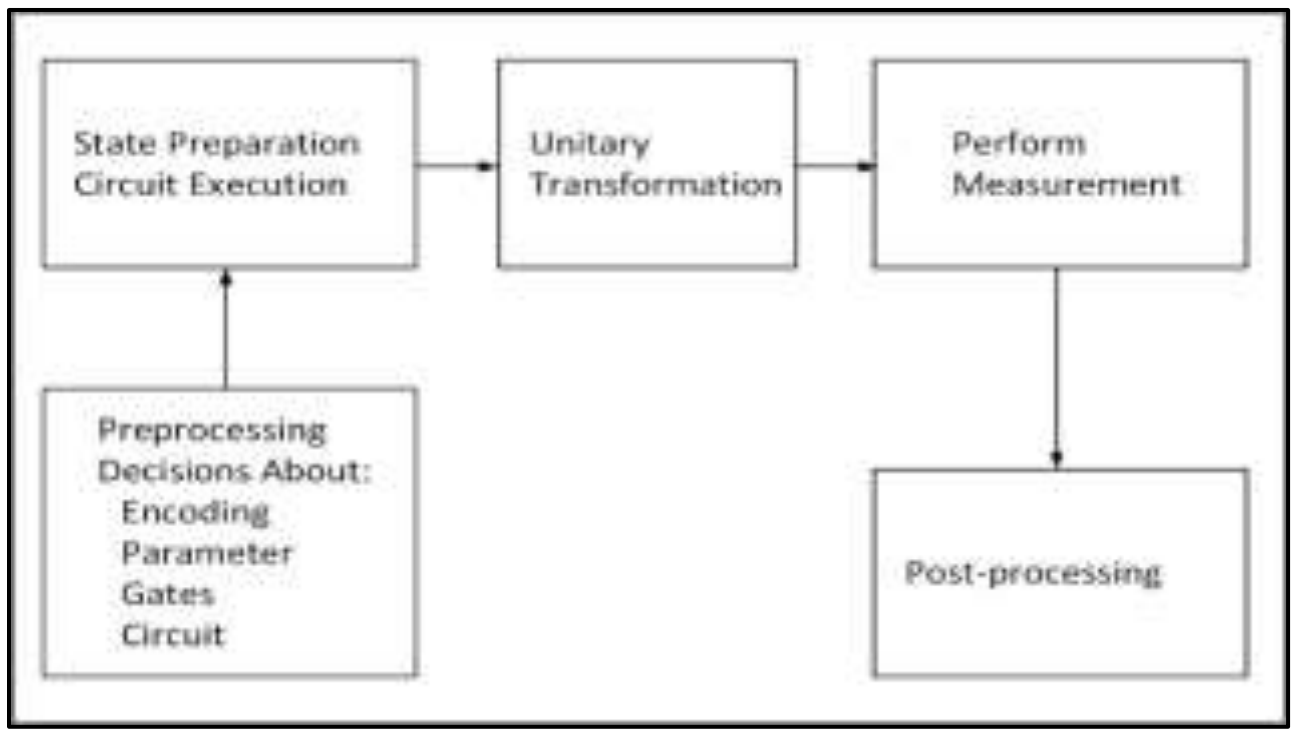

*Fig.1 Classical inputs & quantum algorithms. Adapted from [6]*

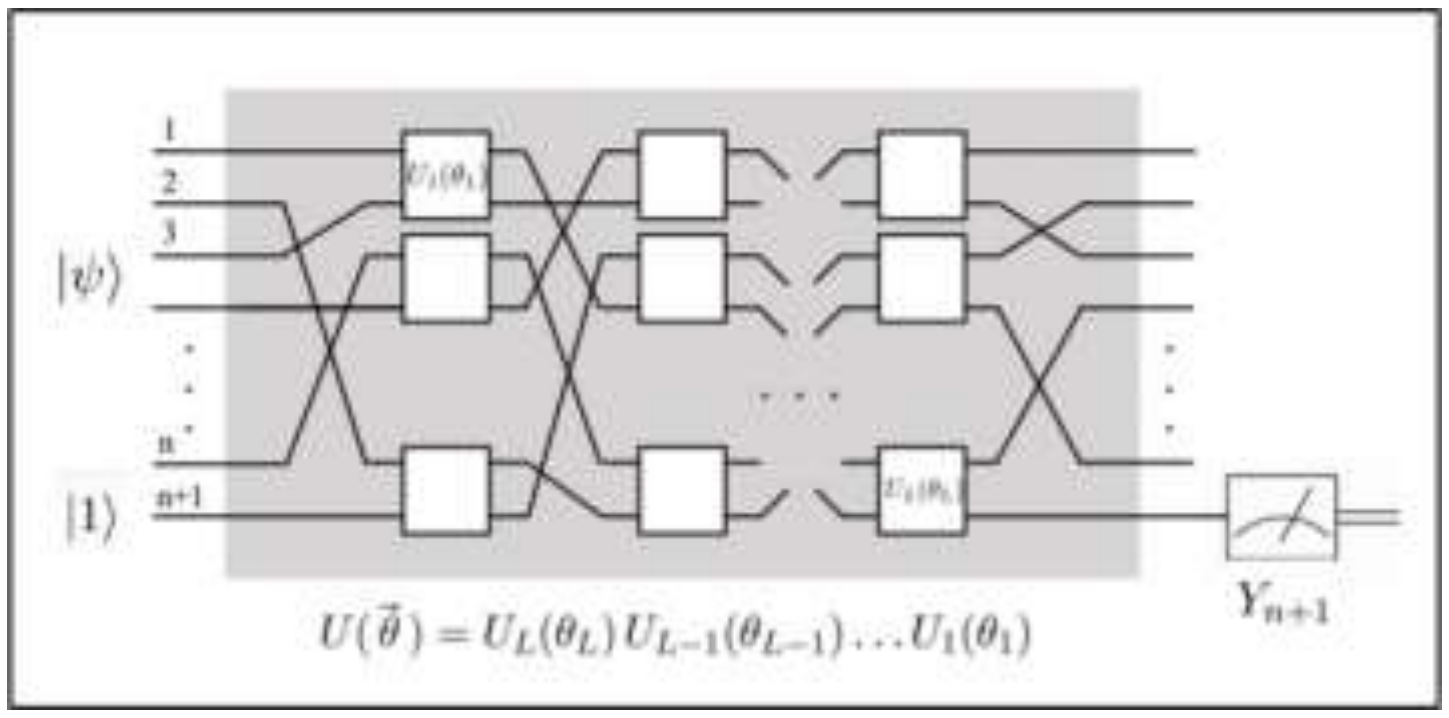

*Fig. 2 QNN on a quantum processor. Source [9]*

In order to solve today's real-life problems, quantum computers need to be able to read, interpret and analyze input datasets. Input data to a quantum system can be any data source that occurs in a natural or artificial quantum system. Information can be encoded into an $n$-qubit system described by a state in various ways. Generally, the method of encoding varies with the nature of the datasets and the problem to be solved. For quantum machine learning and data mining the importance of encoding is paramount. There are quite a few methods of encoding around, for e.g., basis encoding, amplitude encoding, tensor product encoding and Hamiltonian encoding. As per current practice, input preparation has a classical part called preprocessing which creates a circuit that can be processed on a quantum computer to prepare the quantum state. The latter part which actually prepares the quantum state is referred to as state preparation. The process of preprocessing during the classical part in this case may constitute a manual task performed by a human or an automatic task performed by a program. In order to prepare input for a quantum algorithm as a quantum state, a quantum circuit is defined which prepares the corresponding state. This circuit can be generated in a classical preprocessing step as shown in Fig. 1. The generated circuit is prepended to the circuit of the algorithm proper, send to a quantum device and executed. Hence, in order to evaluate the executability of a given algorithm on a particular device, the effort and complexity in terms of additional gates and qubits required to prepare the input of the algorithm proper needs to be considered. The concept of generating the number of gates required to prepare an arbitrary quantum state from classical data is discussed in reference [7]. In this context, efficiency in terms of time and space complexity in encoding classical data into a quantum superposition state suitable to be processed by a quantum algorithm is critical. Formulations of quantum circuits for loading classical data into quantum states for processing by a quantum computer are an active focus of current research [24, 25].

The field of QNN is relatively new as research and work in this field continues to grow. A quantum neural network (QNN) is interpreted today as any quantum circuit with trainable continuous parameters. QNN is a machine learning model that allows quantum computers to classify various datasets and among them, image data [15]. The image data used is classical data, but classical data cannot reach a superposition state. So, in order to carry out this protocol on quantum systems, the data must be made readable into a quantum device that provides superposition.

*Farhi* and *Neven* in 2019 proposed [9] a quantum neural network (QNN), that can represent labeled data, classical or quantum, and be trained by supervised learning. The quantum circuit consists of a sequence of parameter dependent unitary transformations which acts on an input quantum state. For binary classification a single Pauli operator is measured on a designated readout qubit. The measured output is the quantum neural network's predictor of the binary label of the input state. The Reed-Muller representation [10] that is used for subset parity is just another way to think about the two-qubit unitaries. Fig. 2 illustrates a schematic of the quantum neural network proposed by *Farhi* and *Neven* [9] on a quantum processor. $|\psi, \ldots, 1\rangle$ are the input states which are prepared and then fed into the QNN. These input states go through a sequence of *few* qubit unitaries given by $U_i(\theta_i)$ dependent on parameter $\theta_i$ which, in turn, get adjusted during the learning process such that the measurement of $Y_{n+1}$ on the measurement readout tends to produce the desired label for $|\psi\rangle$.

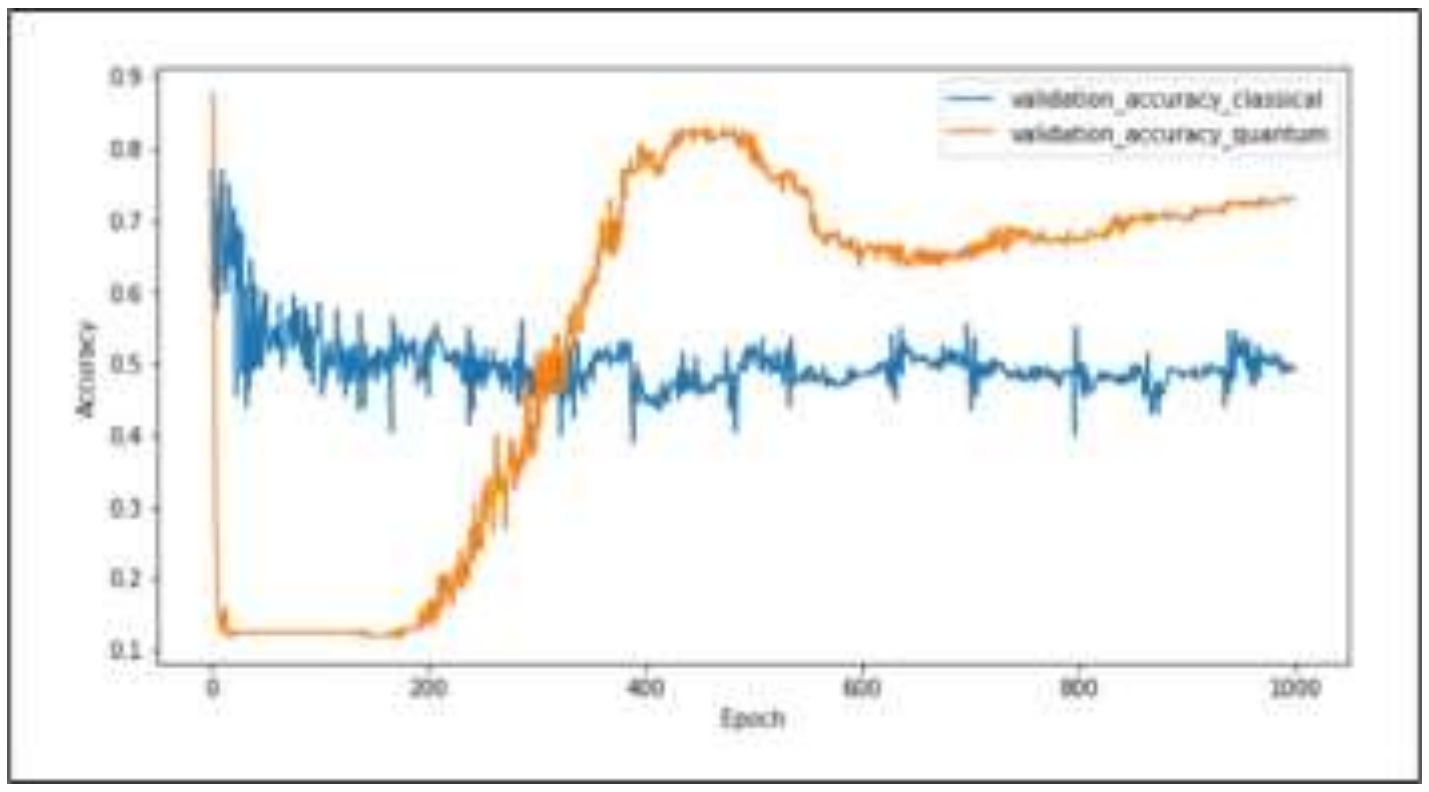

Fig 3. *Comparison between classical (blue) and quantum (orange) NN [26]*

Leveraging the background mentioned above, a previous paper by the authors [26] explored the implementation of QNN to classify image data from Fashion-MNIST, similar to the approach used in *Farhi* et al [9]. Our work [26] showcased an empirical performance separation between classical and quantum machine learning model via a classical image dataset by training a classical neural network on the re-labeled dataset and comparing the performance with a model that had access to the PQK features. We leveraged Principal component analysis (PCA) and Projected Quantum Kernel features (PQK) defined by *Huang* et al. (2021) [11] to study how and when quantum machine learning models can learn as well as (or better than) classical models. Our results leveraging PCA and PQK are summarized in Fig 3, taken from [26].

Classical and Quantum NN models were trained on the Fashion-MNIST data and the performance were plotted for comparison between the two. As shown in Fig. 3. Both models, especially the quantum ones, achieve more than ~90% accuracy on the training data performing better than their classical counterparts. The validation data in Fig. 3 demonstrates the important fact that just the information found in the PQK features is enough to make the model generalize well to unseen instances. Datasets that are interpreted and learnt from relatively easily by quantum models and hard to do so for classical models (for the same datasets) do exist, regardless of model architecture or training algorithms used.

In this study we explored the implementation of QNN to classify image data from Fashion-MNIST. In order to investigate possible improvements in the performance of the QNN, we applied quantum image processing (QIMP) via Novel Enhanced Quantum Representation (NEQR) [25] on the digital image dataset as a pre-processing step and compared that with the classical counterpart. To the best of the knowledge of the authors, no such performance investigation has been done in published work prior to this.

Our results show that machine learning (ML) problems where classical training data is provided can perform better than their quantum counterparts, even when the quantum circuits generating the data are hard to compute classically. We have shown [26] that there are datasets that are interpreted and learnt from relatively easily by quantum models and hard to do so for classical models, regardless of the model architecture or training algorithms used. We ran each neural network model 200 times and accepted the average performance as a benchmark for the model. The QNN showed an average of ~85.0%, the QNN with NEQR showed ~91.0% while the Classical NN easily outperformed both at ~99.85% for this classical image dataset. The application of NEQR did show some nominal improvement ($\sim 5.0\% \pm 1.2$) in performance but still, the classical neural network outperformed for this classical dataset. In addition, NEQR proved to require circuit depths of a proportion that is not realistic in current NISQ era quantum computers. Due to these factors, NEQR may not be an ideal solution for QIMP for classical digital images.

# 2 Method and Results

This section details the approach of this work in terms of dataset, preprocessing of data and results obtained.

## 2.1 Collecting and Processing of Data

Fashion-MNIST[1] [12] is a dataset of Zalando's article images—consisting of a training set of 60,000 examples and a test set of 10,000 examples. Each example is a 28x28 image, associated with a label from 10 classes. Zalando created Fashion-MNIST to serve as a direct drop-in replacement for the original MNIST dataset showcasing handwriting images for benchmarking machine learning algorithms. Fashion-MNIST shares the same image size and structure of training and testing splits as the original MNIST. In Fashion-MNIST, each training and test example is assigned to one of the following labels as shown in Table 1:

**Table 1**

| **Label** | 0 | 1 | 2 | 3 | 4 | 5 | 6 | 7 | 8 | 9 |
|---|---|---|---|---|---|---|---|---|---|---|
| **Item** | T-shirt / Top | Trouser | Pullover | Dress | Coat | Sandal | Shirt | Sneaker | Bag | Ankle Boot |

### *2.1.1 Software*

TensorFlow Quantum[2] (TFQ), proposed by Broughton et al. in 2020 [13] is a Python framework for hybrid quantum-classical machine learning that is primarily focused on modeling quantum data. The relevant code was run on Google Colab utilizing their GPU services – Tesla T4 GPUs were used in this case. Qiskit, run on the QASM [34] simulator, was used to code the NEQR part of this study s currently available quantum hardware cannot realistically support the qubit requirement for NEQR and we prove this in this study. No significant difference was found in terms of performance between Qiskit and Cirq when it came to the NEQR codes. Qiskit was chosen purely because of availability of documentation relating to NEQR [34].

### *2.1.2 Methodology*

The workflow of this paper is shown in Fig. 4. In this work, the classical image data was obtained and downscaled and transformed to quantum data for ease of computing on quantum platforms. The new data was labelled, and the following use cases were investigated: first a performance comparison between classical and quantum neural networks and then NEQR was applied on the same data and fed to the QNN process to investigate performance enhancements if any in the following way: Let $\varphi$ be a set of input data and $\vartheta$ be a feature space. Then a feature map $\Omega$ is defined as $\Omega: \varphi \rightarrow \vartheta$. The output data-points $\varphi(x)$ on the output map are called *feature vectors* and $\vartheta$ is usually a *vector space*.

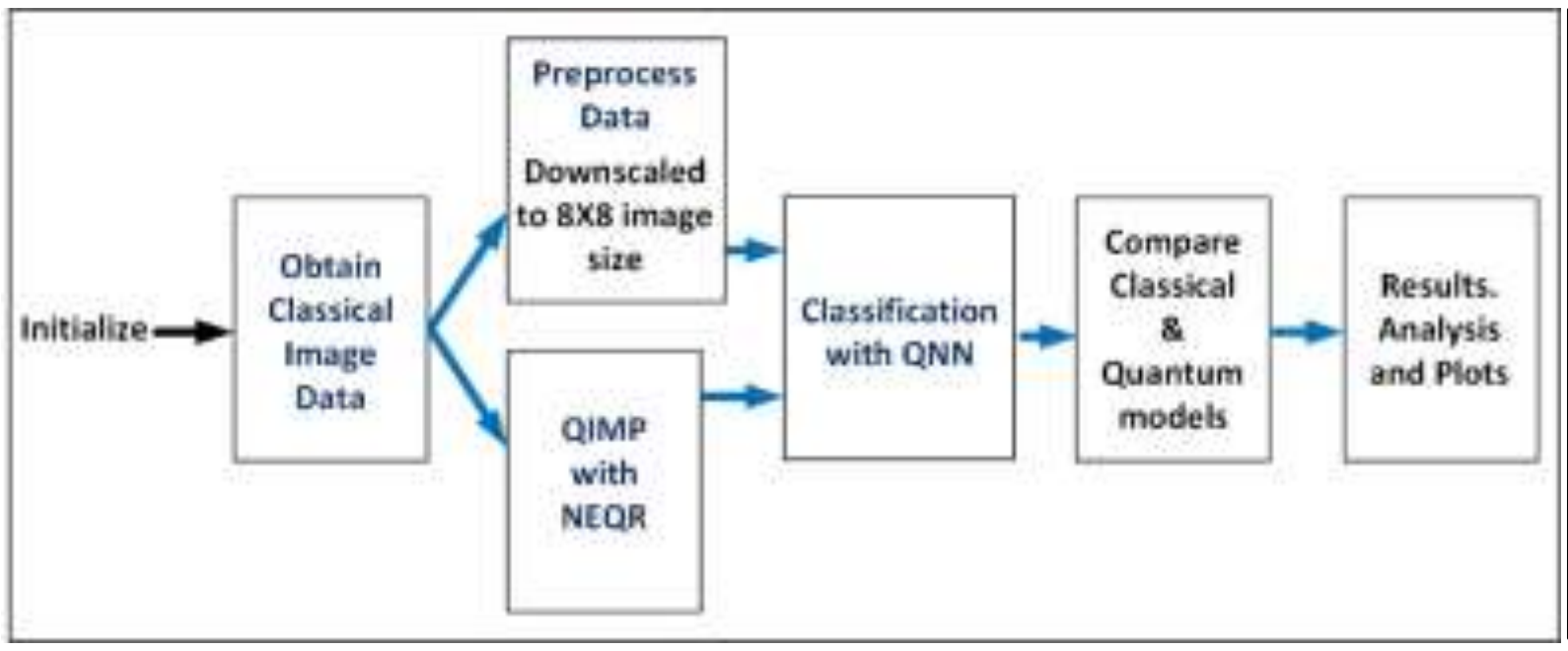

*Fig. 4. Workflow for this study*

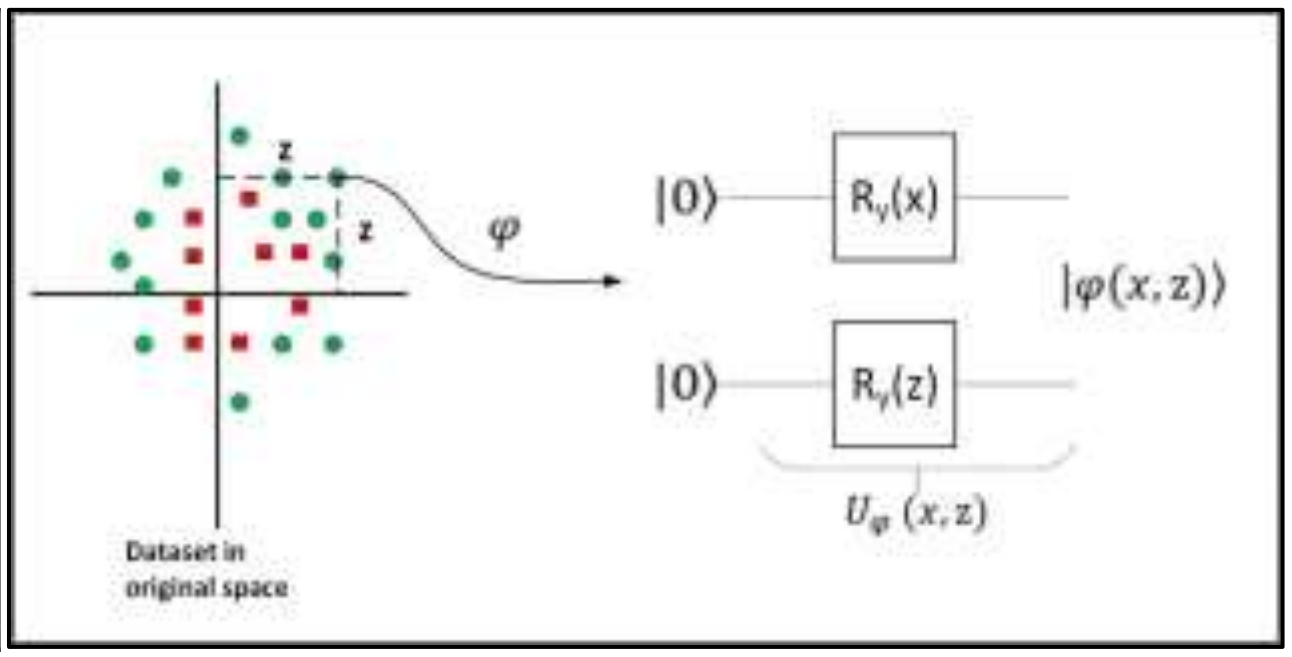

*Fig. 5. QSFT with Variational Circuits*

The transformation relation is shown in Fig. 5. While in the original space of training inputs, data from the two classes 'red squares' and 'green circles' are not separable by a simple linear model (left), we can map them to a higher dimensional feature space where a linear model is indeed sufficient to define a separating hyperplane that acts as a decision boundary (right). For more detail, please see *Schuld* and *Killoran* in ref. [14]. In context of a *quantum feature map*, the feature vectors are quantum states and $\vartheta$ is a member of the Hilbert space. The mapping $\Omega: \varphi \rightarrow \vartheta$ applies a unitary transformation $U$ which, in turn, conducts the following transformation $x \rightarrow |\varphi(x)\rangle$.

[1] Original dataset: https://github.com/zalandoresearch/fashion-mnist
[2] Attribution: "TensorFlow, the TensorFlow logo and any related marks are trademarks of Google Inc."

Typically, $U$, in this case, is a variational circuit and the transformation, known as Quantum Feature Space Transform (QSFT), happens as depicted in Fig. 5. In context of a *quantum feature map,* the feature vectors are quantum states and $\vartheta$ is a member of the Hilbert space. The mapping $\Omega: \varphi \rightarrow \vartheta$ applies a unitary transformation $U$ which, in turn, conducts the following transformation $x \rightarrow |\varphi(x)\rangle$. Typically, $U$, in this case, is a variational circuit and the transformation happens as depicted in Fig.5.

#### *2.1.3 Quantum Embedding*

Generally speaking, a quantum algorithm is fed input data that must be represented as a quantum state in order to be understood by and manipulated on a quantum computer. Most applications for QML today uses quantum feature maps to map classical data to quantum states in a Hilbert space. This is a very important aspect of designing quantum algorithms which has a direct impact on their computational cost. The process is known as *quantum embedding* and involves translating a classical datapoint $x$ into a set of gate parameters in a quantum circuit, creating a quantum state $|\psi_x\rangle$. More details can be found in [16] and [17]. If we consider classical input data consisting of $P$ datapoints, with $R$ features each, then, for a dataset $D$,

$$D = x^1, x^2, \dots, x^p, \dots, x^P$$

where, $x^p$ is a $R$-dimensional vector and $p = 1,2, \dots, P$. This data can be embedded or encoded into $n$ quantum *subsystems*. These subsystems can be $n$-qubits for discrete-variable (DV) or $n$-qumodes for continuous-variable (CV) quantum computing. Determining which features capture the *largest variance* is the goal of *Principal Component Analysis* (PCA) [18]. Mathematically, PCA involves taking the raw data (e.g., the feature vectors for various houses) and computing the covariance matrix.

#### *2.1.4 QIMP with NEQR*

Quantum image processing (QIMP), which utilizes the characteristic of quantum parallelism to speed up many processing tasks, is a subfield of quantum information processing [3, 27]. Research has shown that quantum computers can utilize maximally entangled qubits to reconstruct images without additional information [31], improving both storage and retrieval. Using a quantum measurement to probe the entanglement shared between the vertex qubits can be used to determine their location. Furthermore, the parallelism that is inherent in quantum systems have been used in fast image searching [33] in addition to image reconstruction. *Image segmentation*, which is the process of dividing an image into separate regions (for example finding faces in a picture), is an important aspect of image processing in quantum computation and the concept is especially important for machine learning when the detection of this region is automated. *Venegas-Andraca* et al. [31] expands on these techniques and how this improves image segmentation relative to traditional systems.

Various representations for images on quantum computers have been proposed, such as qubit lattice, wherein the images are two-dimensional arrays of qubits [32]; flexible representation of quantum image (FRQI), wherein the images are normalized states that capture the essential information about every point in an image, i.e., its color and position [27, 28, 29]; multi-channel quantum image (MCQI) [30], which is an extension of FRQI representation that contains the R, G and B channels for processing color information and Novel Enhanced Quantum Representation of digital images (NEQR) which works with an internal representation of an image [25]. The representation of a $2^n \times 2^n$ NEQR image is given by [25] as,

$$|I\rangle = \frac{1}{2^n} \sum_{y=0}^{2^n-1} \sum_{x=0}^{2^n-1} |f(y,x)\rangle\, |yx\rangle = \frac{1}{2^n} \sum_{y=0}^{2^n-1} \sum_{x=0}^{2^n-1} \bigotimes_{i=1}^{q-1} |C_{yx}^i\rangle\, |yx\rangle$$

where the grayscale value $f(y,x) = C_{yx}^{q-1} C_{yx}^{q-2} \dots C_{yx}^1 C_{yx}^0$ and $C_{yx}^i \in [0,1]$, $f(y,x) \in [0, 2q-1]$. In this paper NEQR was used for QIMP. NEQR uses the basis state of a qubit sequence to store the grayscale value of every pixel. Therefore, two qubit sequences, representing the grayscale and positional information of all of the pixels, are used in NEQR representation to store the whole image. The computational complexity of preparing an NEQR image shows a quadratic decrease, i.e., $O(qn.2^n)$, compared to FRQI and MCQI images [25]. However, *it is to be noted* that NEQR representation uses *more qubits* and requires higher circuit depths to encode a quantum image as we shall demonstrate. From its representation, $(q + 2n)$ qubits are needed to construct the quantum image model for a $2^n \times 2^n$ image with gray range $2^q$. However, the $2n$ qubits for position information is the same as for FRQI and MCQI representation. NEQR uses $q$ qubits for color information, while FRQI and MCQI use 1-qubit and 3-qubits, respectively. NEQR uses normalized superposition to store pixels in an image and was created to improve over FRQI by leveraging the basis state of a qubit sequence to store the grayscale value of an image.

### 2.2 QNN and Classical NN

The Fashion-MNIST dataset used to train and test the models was obtained from TensorFlow's `tf.keras.datasets` module. They were split up in train and test samples of 60000 and 10000 respectively. Subsequently, the dataset was sorted and filtered to keep just the T-shirts, tops and dresses and remove all the other classes. The label, $y$, was converted to boolean: 0 for “True” and 3 for “False”. This step provided 12000 training examples and 2000 test samples. In order to verify whether the data was sorted correctly a random training data image was printed as shown in Fig. 6.

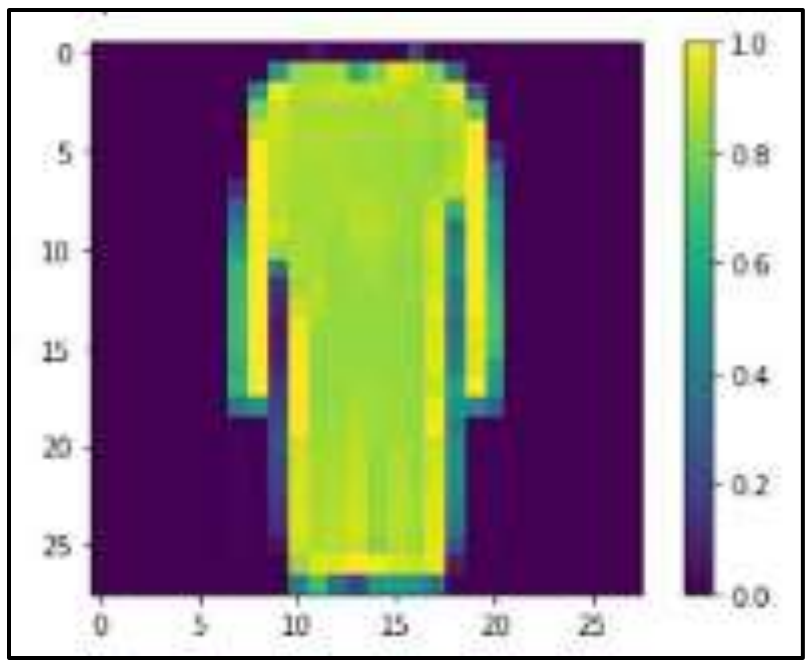

*Fig. 6. Item under label 7 post filtering*

The next step involves downscaling of the dataset. The classic Fashion-MNIST image data size is 28x28 pixels which is too large for currently available quantum computers. Hence, the images were resized to a dimension of 10 using a PCA transformation so that it can be fed into the quantum circuit. As an initial trial run, the size of the datasets was reduced to a training datapoint of 1000 and a test datapoint of 200.

Encode the Data: In the next step the quantum circuit to encode the data for the QNN model is created. An image is made up of pixels. Each pixel in the classic image data was defined to represent a qubit as per [9] as a preparation of quantum data from classical counterparts. The qubits at pixel indices were rotated through a $CNOT$ or $X$ gate and the corresponding circuits were created. Fig. 10 shows an example of such a circuit.

## 2.3 QNN with NEQR and Classical Image Data

In this study we followed the process of Fig 4b and investigated the effect of QIMP with NEQR on QNN using the same Fashion-MNIST image dataset.

### *2.3.1 QNN with Classical Image Data*

In this case the Fashion-MNIST image data was first resized down to $8 \times 8$ size from $28 \times 28$ for ease of computing. *Farhi* et al. [9] proposed representing each pixel with a qubit in order to process images using a quantum computer, with the state depending on the value of the pixel. Hence, the first step was to convert to a binary encoding and the qubits at pixel indices with values exceeding a threshold value of 0.5 were rotated through an $X$-gate and the respective circuits produced – as an example, the first circuit is shown in Fig 7. The full circuit and associated NEQR code is shown in Appendix A.

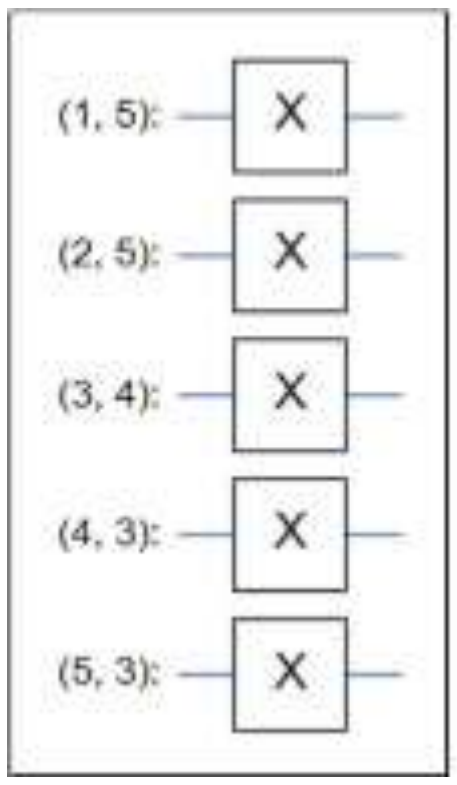

*Fig.7. Circuit*

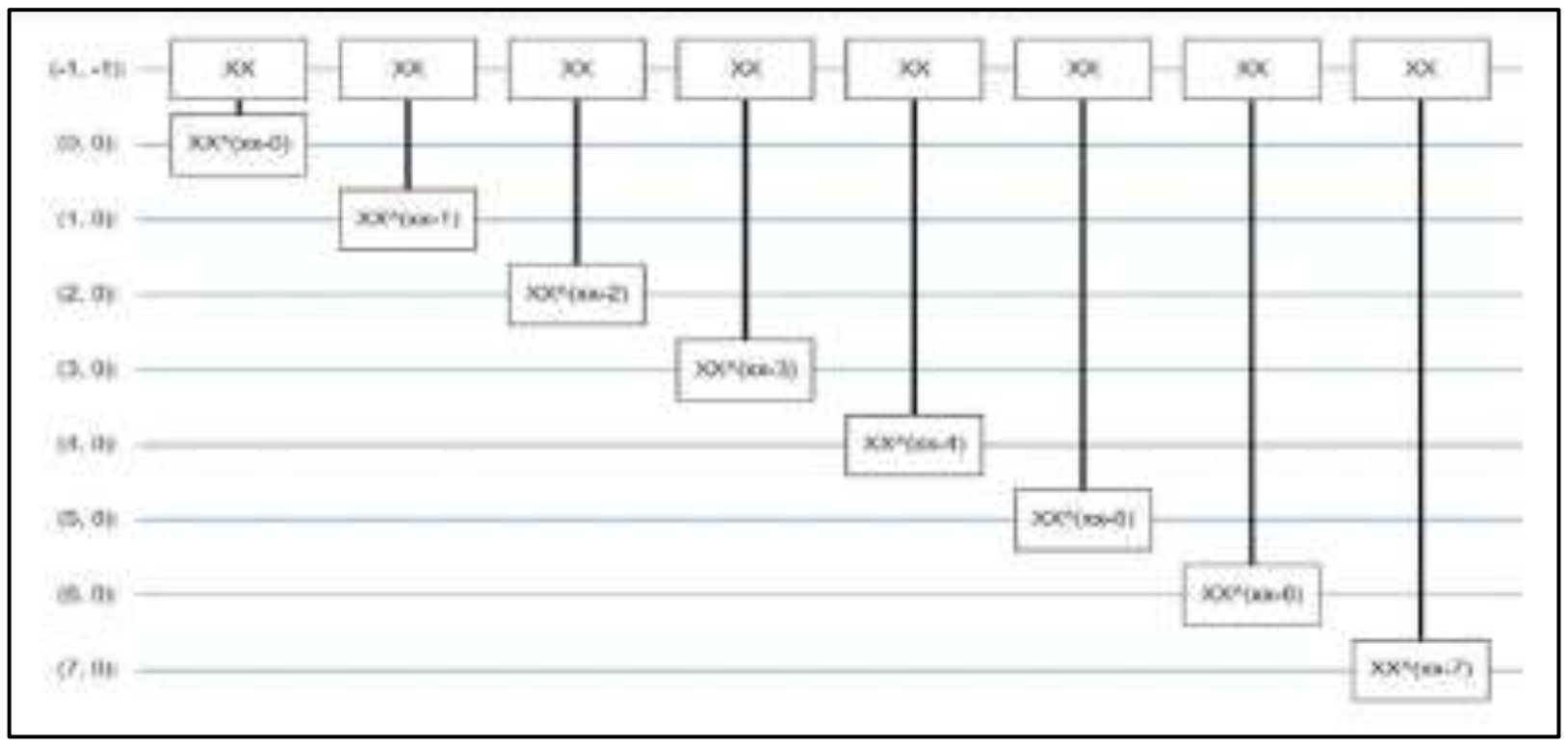
*Fig. 8. A Circuit Layer for the QNN*

In the next step, these circuits, created in `cirq`, were converted to tensors via the `tfq.convert_to_tensor()` function of TensorFlow Quantum. To create the QNN, a process was implemented where the readout qubits were defined to be acted upon by two qubit gates following the suggestion in [9] which is analogous to running small a Unitary RNN across the pixels. To build the QNN model utilizing the above theory, a layered approach was used where each layer had $n$ instances of the same gate, with each of the data qubits acting on the readout qubit. An example circuit layer of this QNN is shown in Fig 8. Next, a two-layered model was built matching the data-circuit size which included the readout operations. A Keras model was built with the quantum components and fed the "quantum data" that *encodes* the classical data. A Parametrized Quantum Circuit (PQC) layer, `tfq.layers.PQC`, to train the model circuit on the quantum data. The readout range, as defined, is $[-1,1]$ which indicates that optimization of the hinge loss should be a natural fit. In order to use the hinge loss as a parameter, two adjustments were made: the labels were converted from boolean to $[-1,1]$, and a custiom `hinge_accuracy` metric was introduced to handle $[-1,1]$ as the `y_true` labels argument. The output of the model summary is shown in Fig. 9 below. In the next step the quantum model was trained for 20 epochs which took approximately 54 min 26 sec on Google Colab with Tesla T4 GPU support. We had a hinge accuracy of ~85%.

### *2.3.2 Classical NN with Classical Image Data*

In the next step we took a look at the performance of a classical neural network for this same reduced Fashion-MNIST dataset. We expected a classical neural network to achieve more accuracy on the holdout set for this particular dataset. The `sigmoid` activation function was used in conjunction with the `Adam` optimizer in this case and the model easily converges to ~99.9% accuracy of the test set.

### *2.3.3 NEQR processed Classical Image Data*

In the last section of this study, we created the NEQR code and processed the classical images as per the workflow of Fig. 4. This is done because in order to be able to apply quantum computing algorithms to process an image, the image information should be stored quantum data format. The workflow of the NEQR model is shown in Fig. 9.

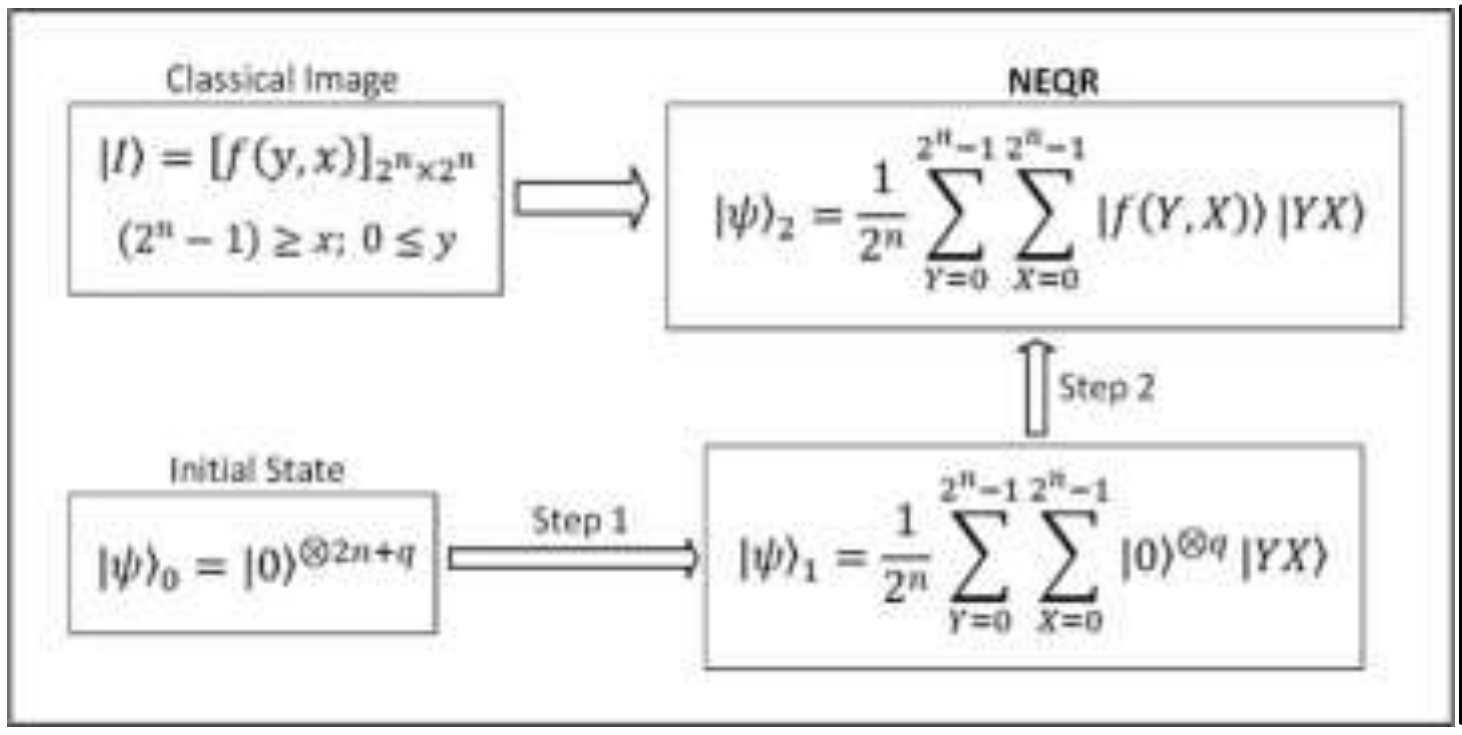

*Fig. 9. Workflow of NEQR image preparation. Adapted from [25]*

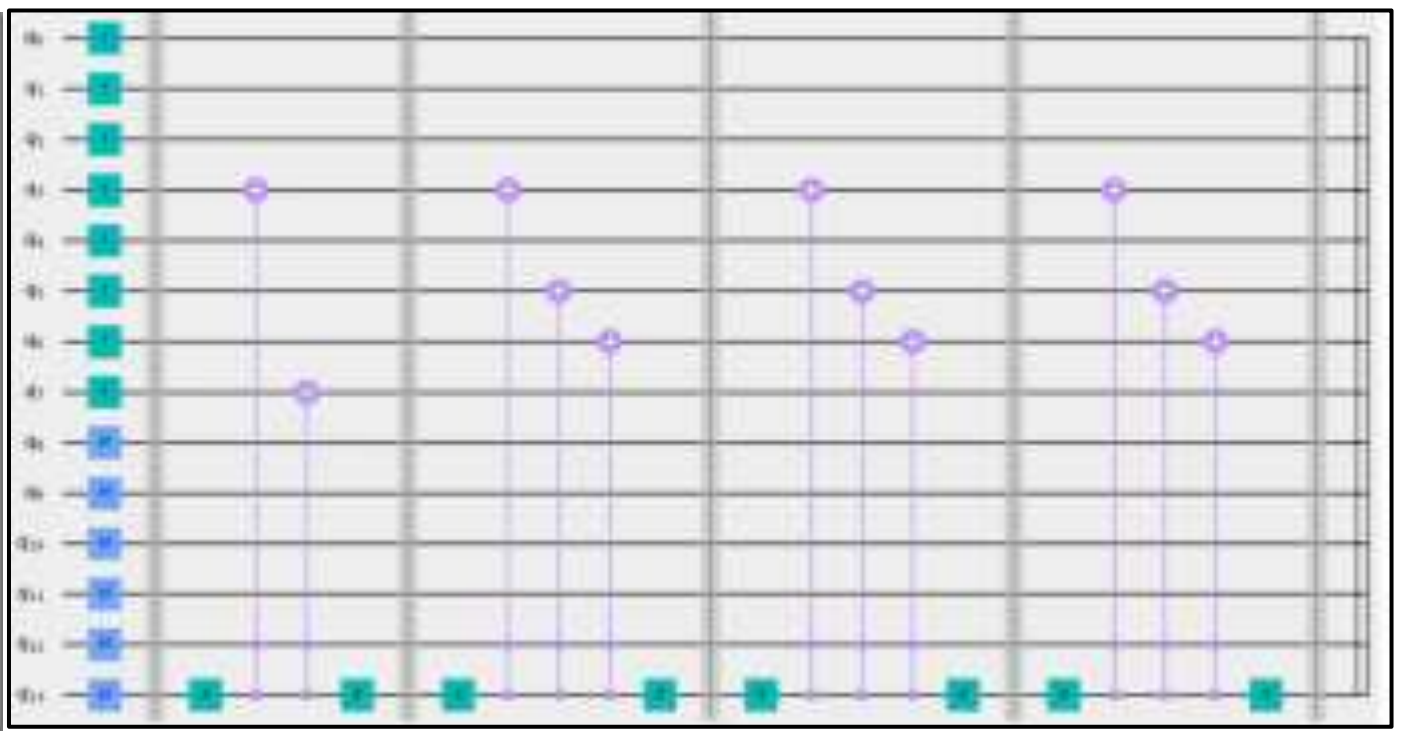

*Fig. 10. A very small snapshot of the NEQR circuit.*

In Fig. 9, the initial state $|\psi\rangle_0 = |0\rangle^{\otimes 2n+q}$is the state after $(2n + q)$ qubits are prepared and set to state $|0\rangle$ which is then transformed into the NEQR model in two steps: preparation and compression. The preparation steps followed were the same described theoretically in [25]. In [25] the authors have not mentioned the type of system, code or platform that they may had used to get confirmation of their theories. In our case, we used QASM simulator from Qiskit for the NEQR part and Cirq and TensorFlow Quantum for the quantum and classical neural network analysis. First, a quantum circuit was created for the pixel values and then another circuit was created for the pixel positions. To create the first circuit, the range of the intensity for each pixel was defined: 8 qubits were required since $256 = 2^q$, where the range of the pixels are $0 - 256$. The images encoded were reduced to $8 \times 8$ due to computational cost. This meant that the second quantum circuit needed 6 qubits to represent the pixel positions since we represented an $8 \times 8$ image with 64 positions and $64 = 2^6$. Therefore, the final circuit was built with $8 + 6 = 14$ qubits.

### *2.3.4 Model Comparison*

A small snapshot of generated circuit, depth and gate decomposition is shown in Fig 10 as the real circuit depth was 2319 with a circuit size of 3427 which is much larger than what a real NISQ era present-day quantum computing platform can support. With this in mind, we proceeded to investigate the benefits of NEQR in terms of performance as compared to our results with QNN and classical NN performed before [26] and obtained the efficiency for a batch run of 200 as shown in Fig 11. As shown in Fig. 11, the classical neural network (NN) model outperforms the QNN counterpart with higher resolution input for this specific dataset. The classical model of similar power trains to a better accuracy in a fraction of the time. Our results showed marginal improvements (only about ~5.0%) where the QNN performance with NEQR exceeded the performance of QNN without NEQR.

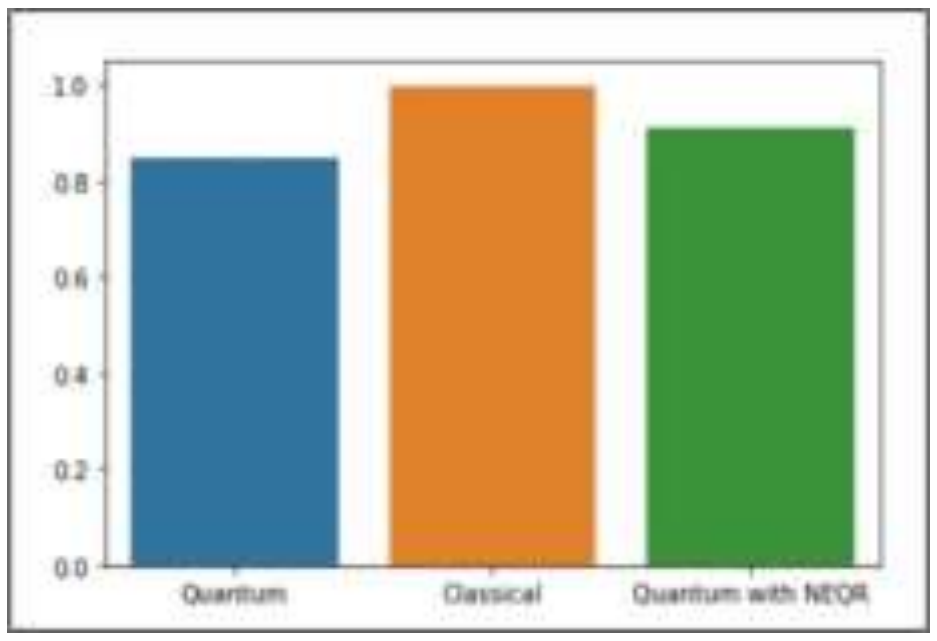

*Fig. 11. Classical NN, QNN & QNN with NEQR*

# 3 Conclusion

In summary, we arrived at several conclusions: a) Performance of the quantum models available today appear to be not always more efficient than their classical counterparts when *classical* data is used. b) The data may not be hard for a classical model to learn from even though it may have come from a hard to classically simulate quantum circuit. c) Datasets that are interpreted and learnt from relatively easily by quantum models and hard to do so for classical models (for the same datasets) do exist, regardless of model architecture or training algorithms used. d) We conclude that given the computational cost and the massive circuit depth associated with running NEQR, the advantage offered by the QIMP algorithm is questionable for classical image dataset. No actual quantum computing hardware platform exists today that can support the circuit depth needed to run NEQR even for the reduced image sizes of our toy classical dataset. Our results showed marginal improvements ($\sim 5.0\% \pm 1.2$) where the QNN performance with NEQR exceeded the performance of QNN without NEQR. As a next step in our continuing studies we intend to investigate a performance analysis of *Quantum Boolean Image Processing* (QBIP).

# 4 Acknowledgment

The author acknowledges the time given by Cisco Systems to work on this side project.

# Appendix A

## Code: Listing A1

```
final_output = []

#for 100 images of the toy dataset

for j in range(100):
  num_qubits = 8 + 6 #8 qubits for pixels and 6 qubits for data
  qc_image = QuantumCircuit(num_qubits)

  # Create the pixel position qubits, and place them in superposition.
  qc_pos = QuantumCircuit(6)
  qc_pos.h(0)
  qc_pos.h(1)
  qc_pos.h(2)
  qc_pos.h(3)
  qc_pos.h(4)
  qc_pos.h(5)

  # Setup color value qubits
  qc_grayscale = QuantumCircuit(8)
  for idx in range(8):
      qc_grayscale.i(idx)

  # Compose the circuit by appending the pixel and grayscale qubits
  qc_image.compose(qc_pos, qubits=[num_qubits-1,num_qubits-2,num_qubits-
3,num_qubits-4,num_qubits-5,num_qubits-6], inplace=True)
  qc_image.compose(qc_grayscale, qubits=[0, 1, 2, 3, 4, 5, 6, 7], inplace=True)
  qc_image.barrier()

  # Add the CNOT gates
  for i in range(64):
    qc_image.x(num_qubits-1)
    for idx, px_value in enumerate(data[j * 64 + i]):
        if(px_value=='1'):

            qc_image.ccx(num_qubits-1,num_qubits-2, idx)
    qc_image.x(num_qubits-1)
```

```
    qc_image.barrier()

  #run circuit in backend and get the state vector
  backend = Aer.get_backend('statevector_simulator')
  result = qiskit.execute(qc_image, backend=backend).result()
  output = result.get_statevector(qc_image)
  final_output.append(output)

  #draw the last circuit
from qiskit.tools.visualization import circuit_drawer
qc_image.draw()
circuit_drawer(qc_image, filename='./qc01', output='mpl', style={'backgroundcol
or': '#EEEEEE'})
circuit_drawer(qc_image, filename='./qc02', output='text', style={'backgroundco
lor': '#EEEEEE'})
```

*Fig. A1.Code to generate Circuit for the NEQR process*

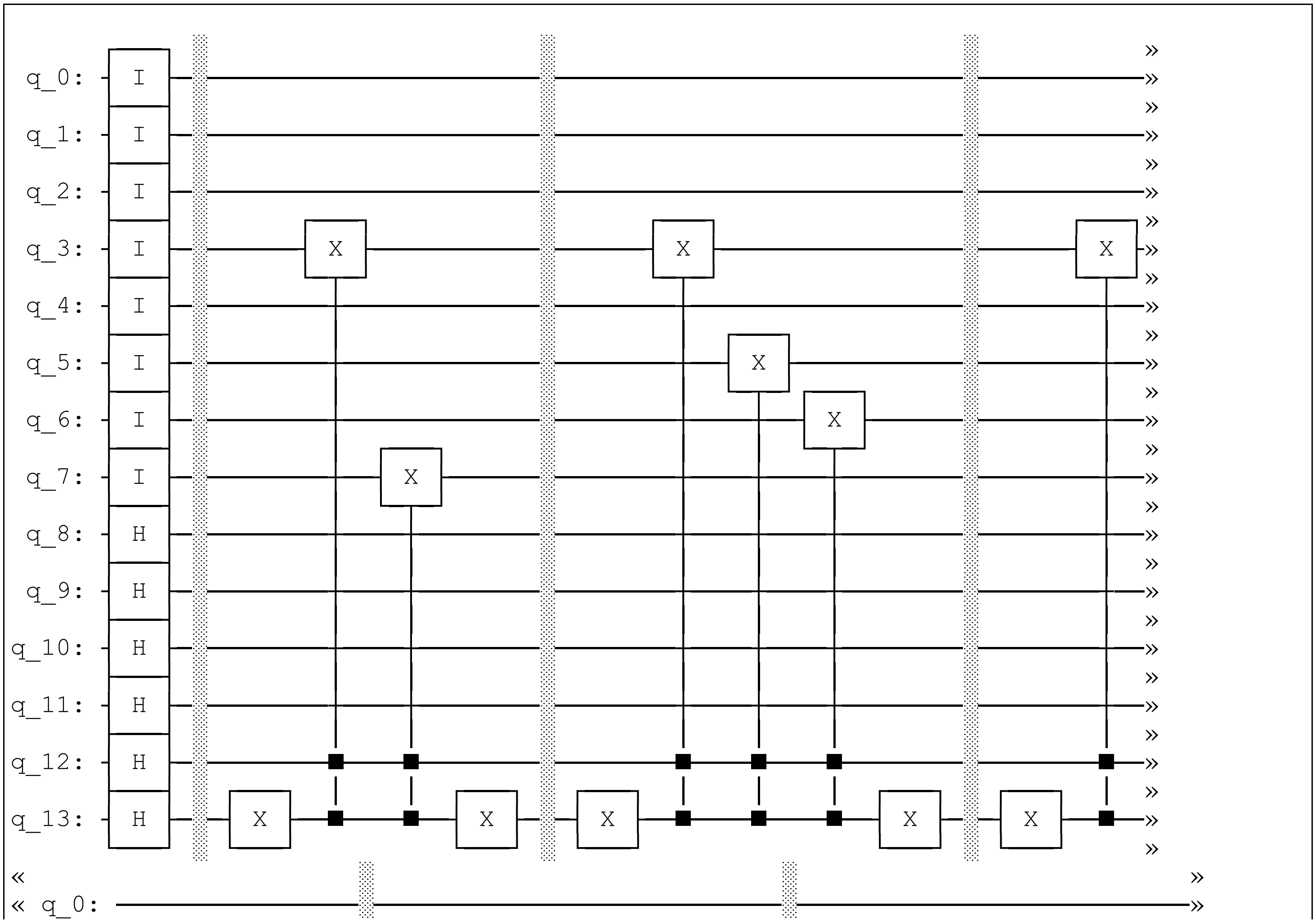

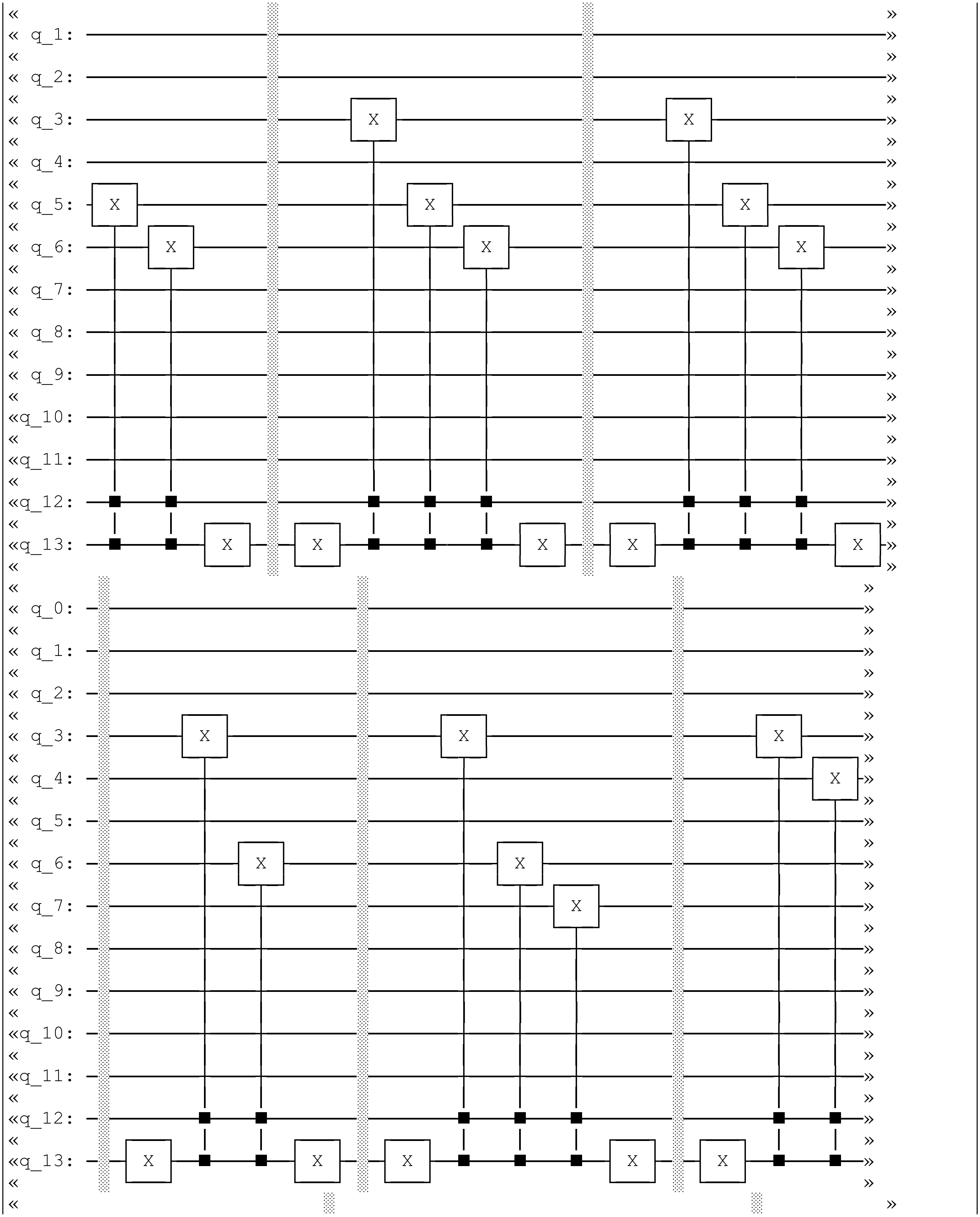

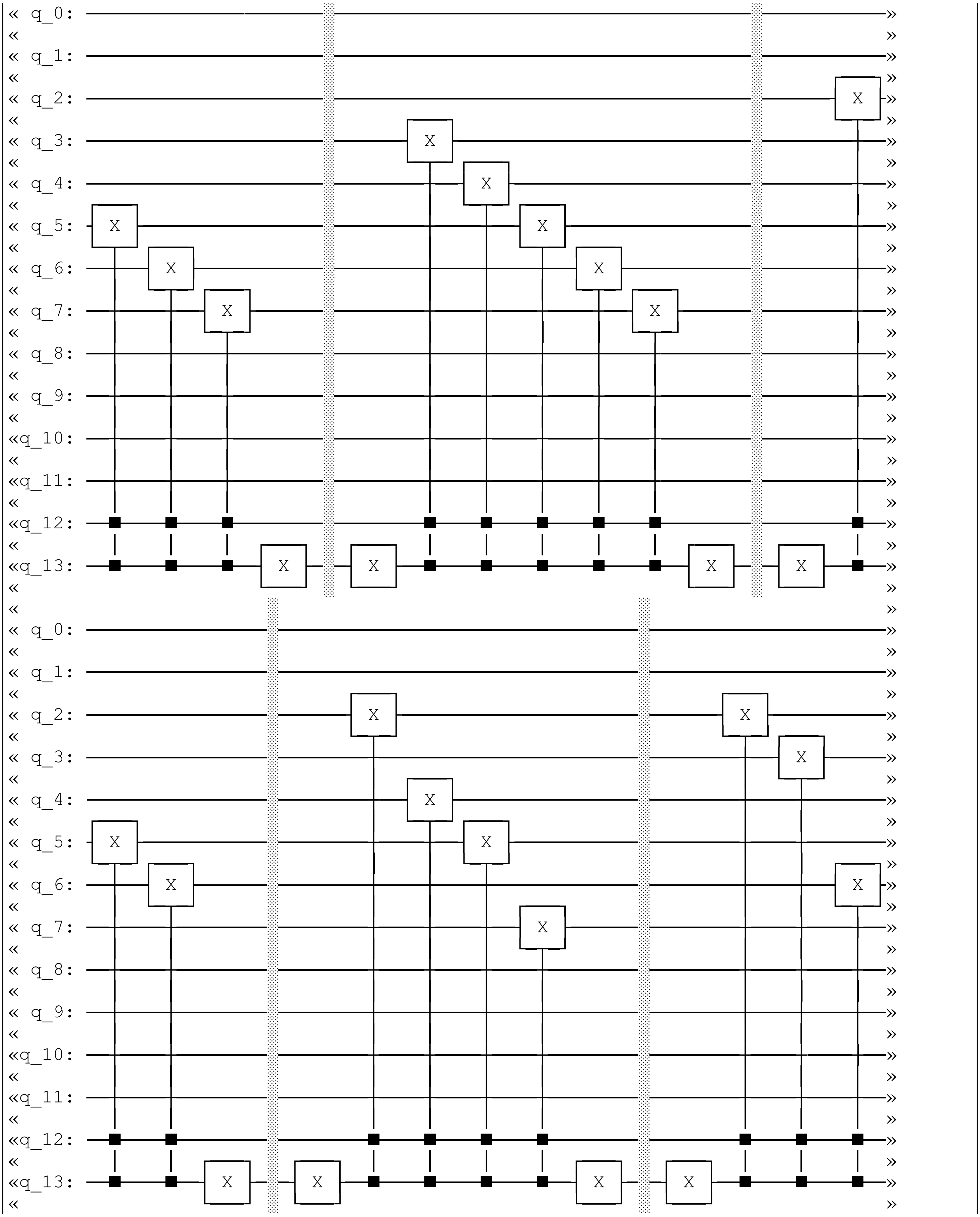

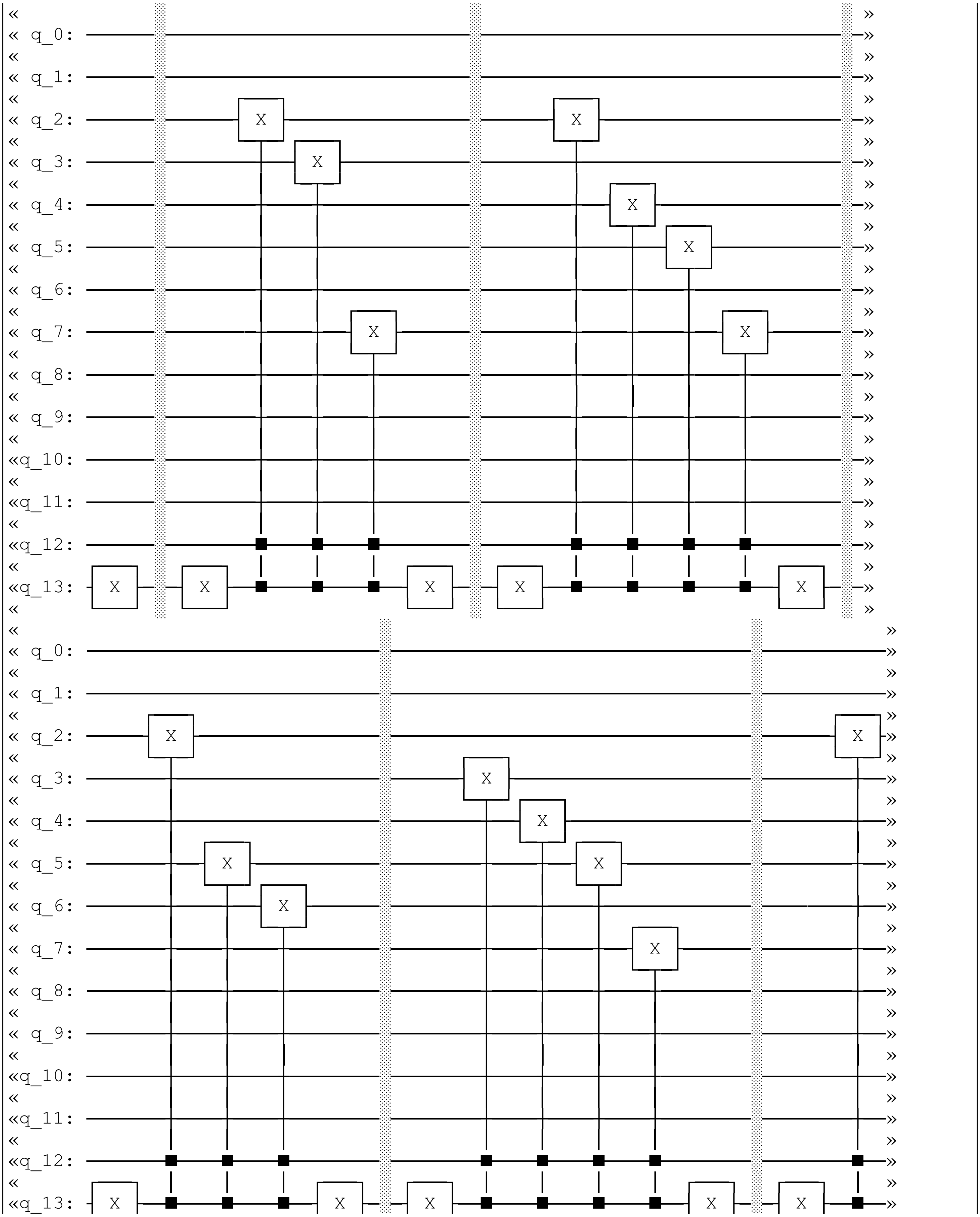

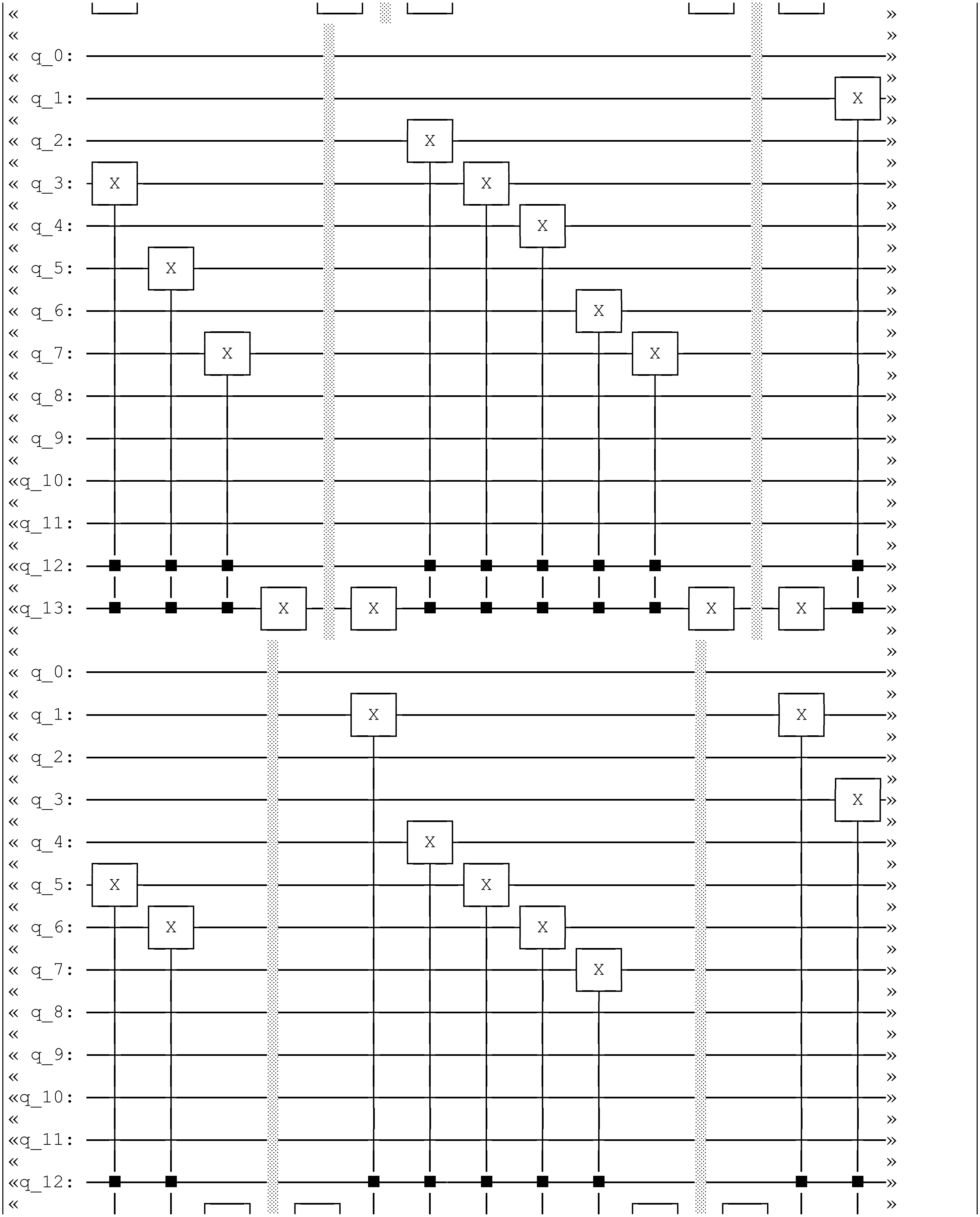

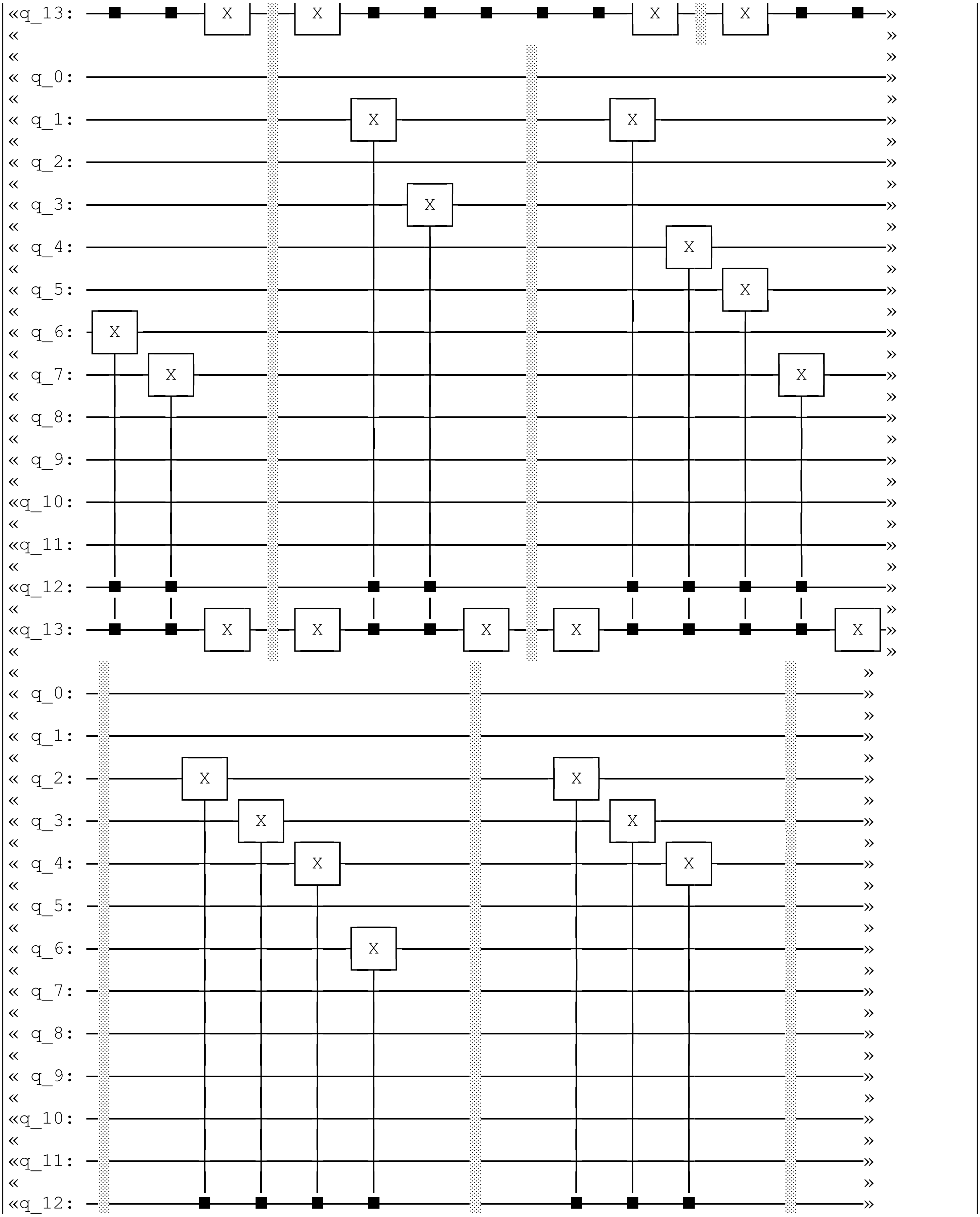

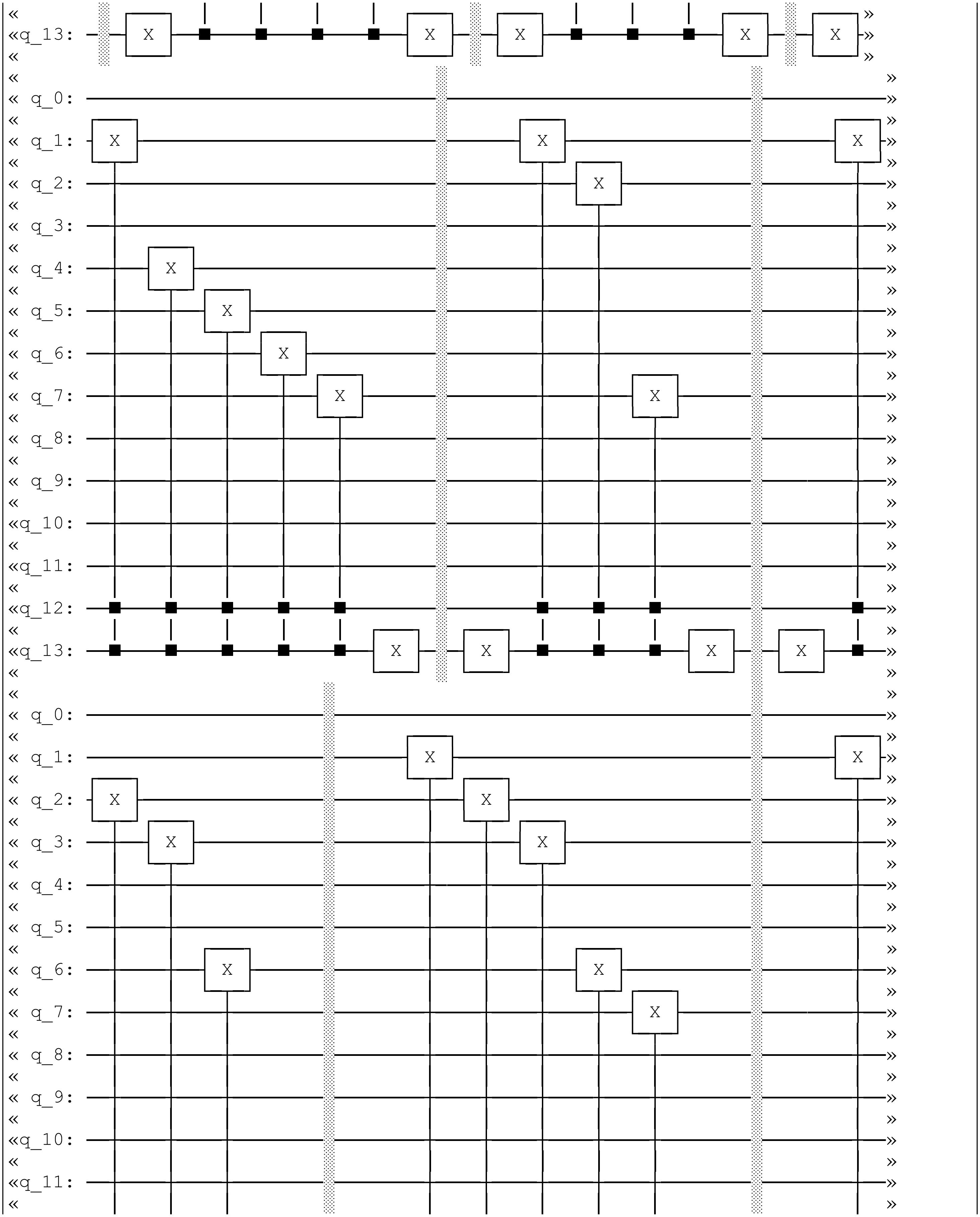

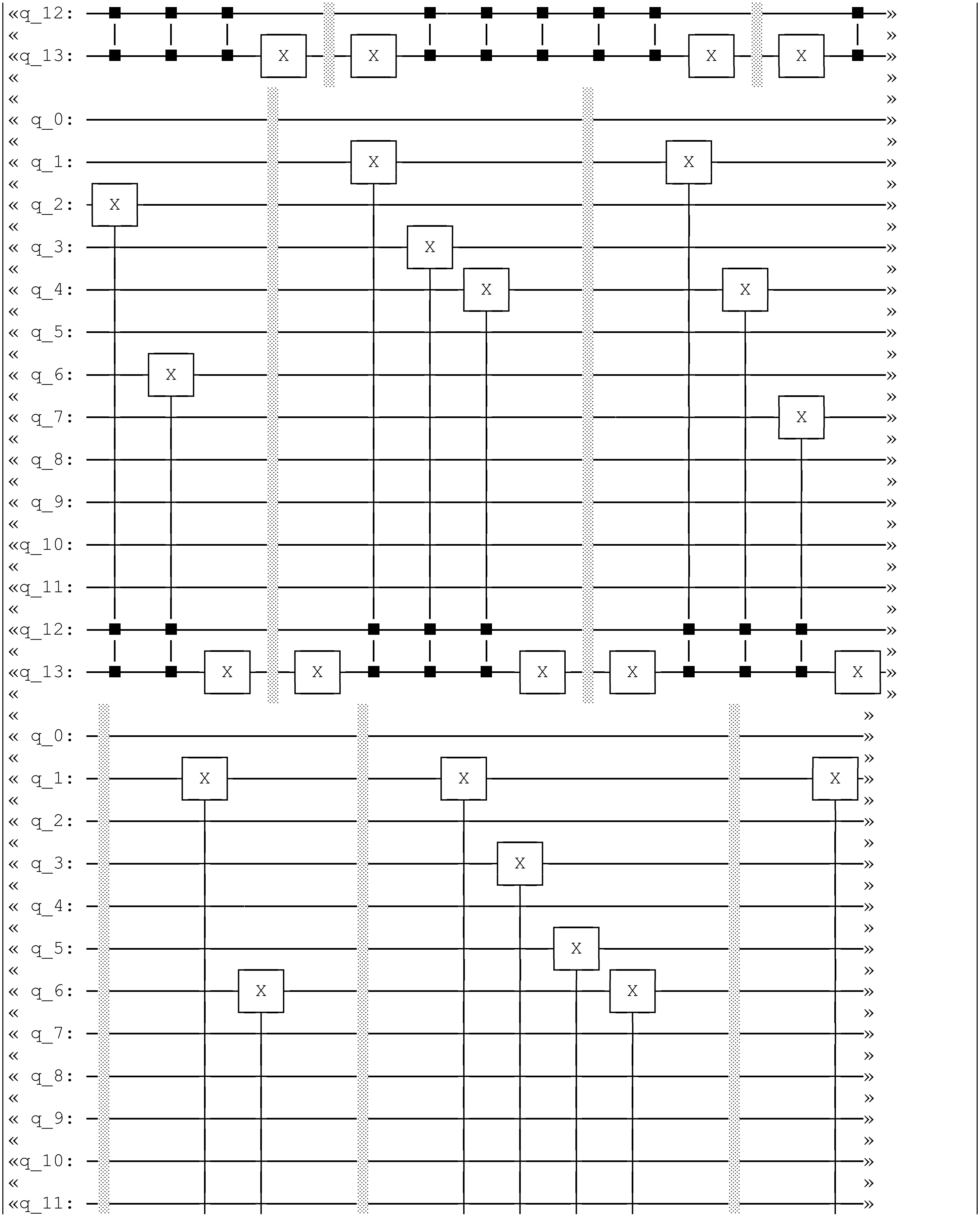

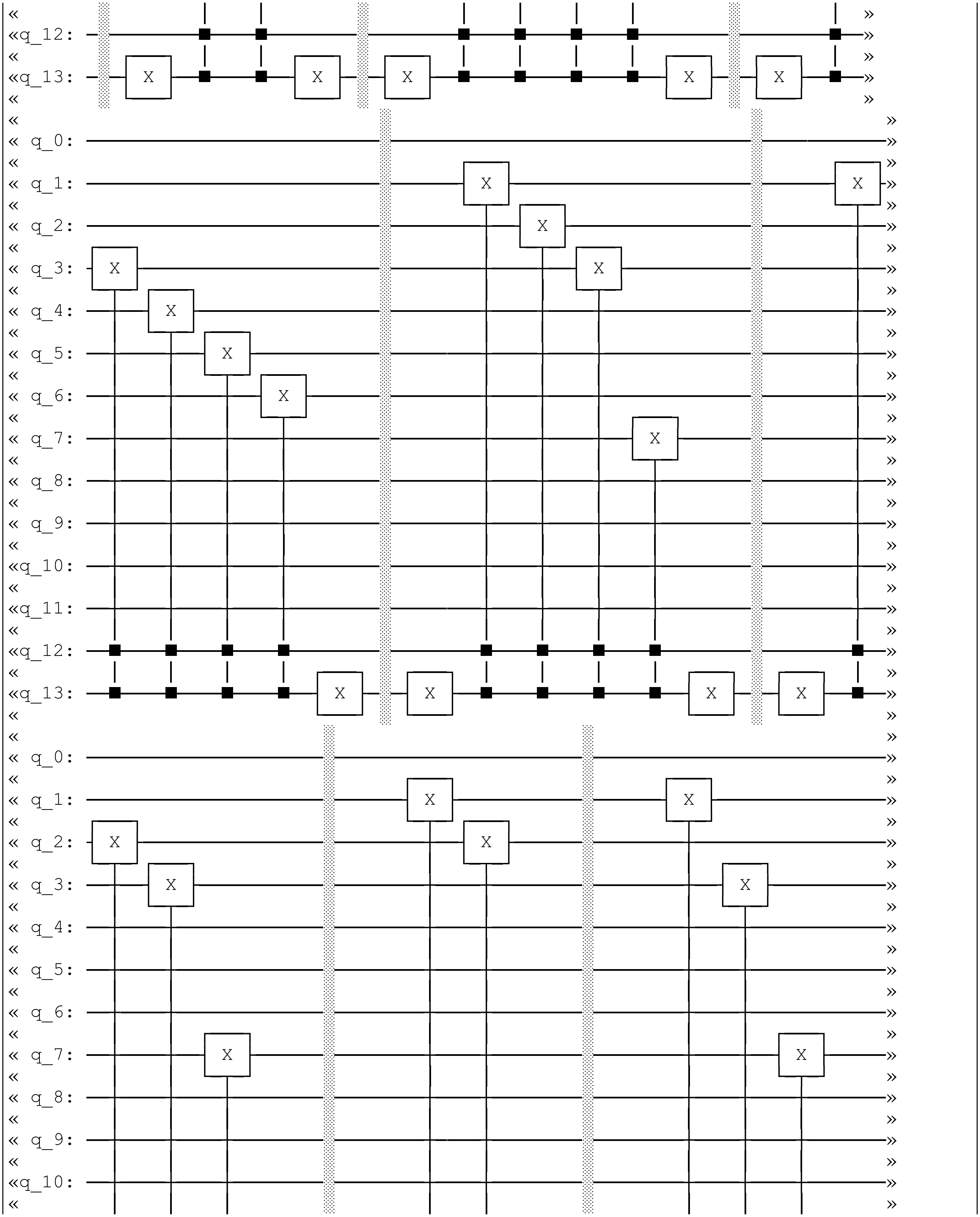

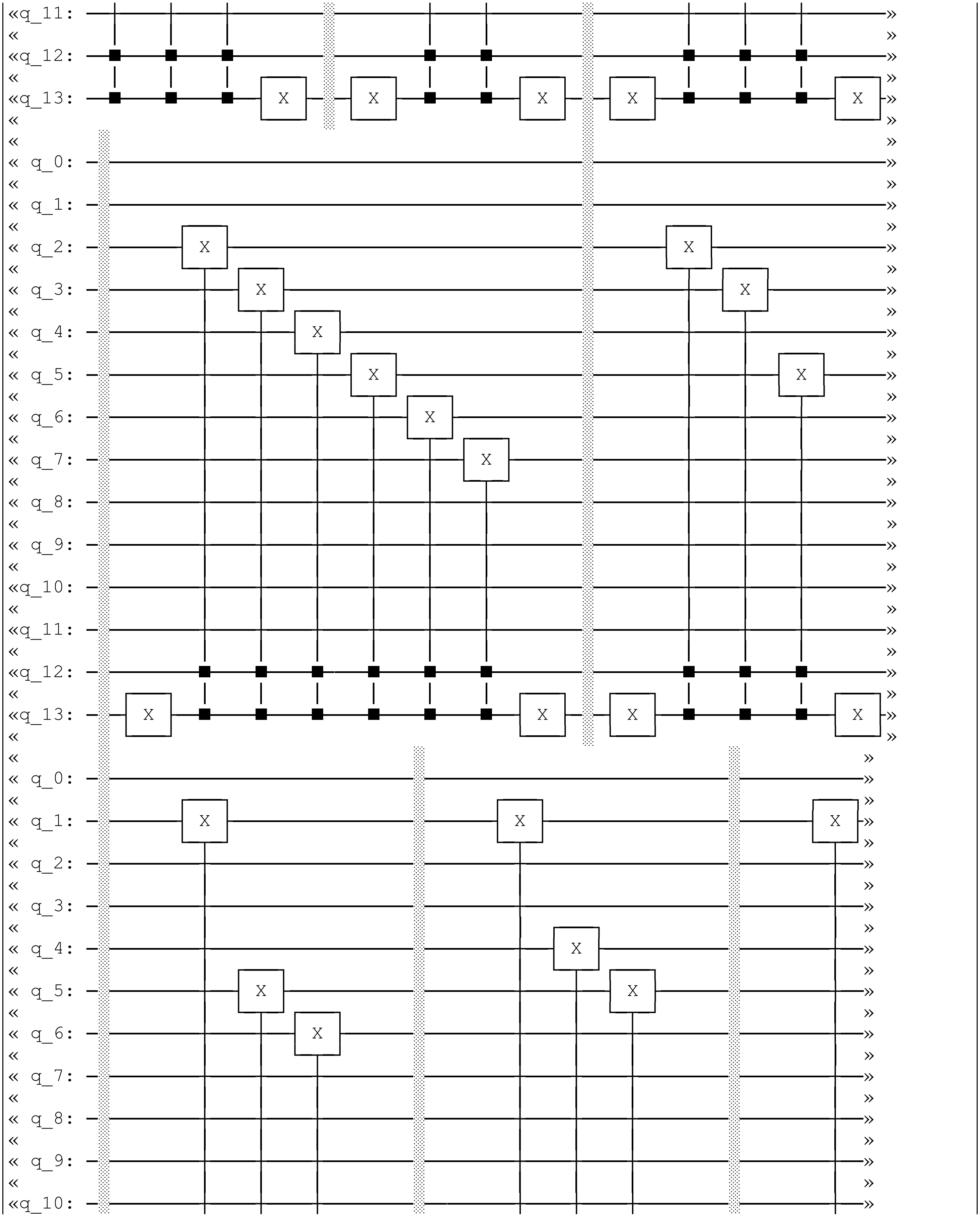

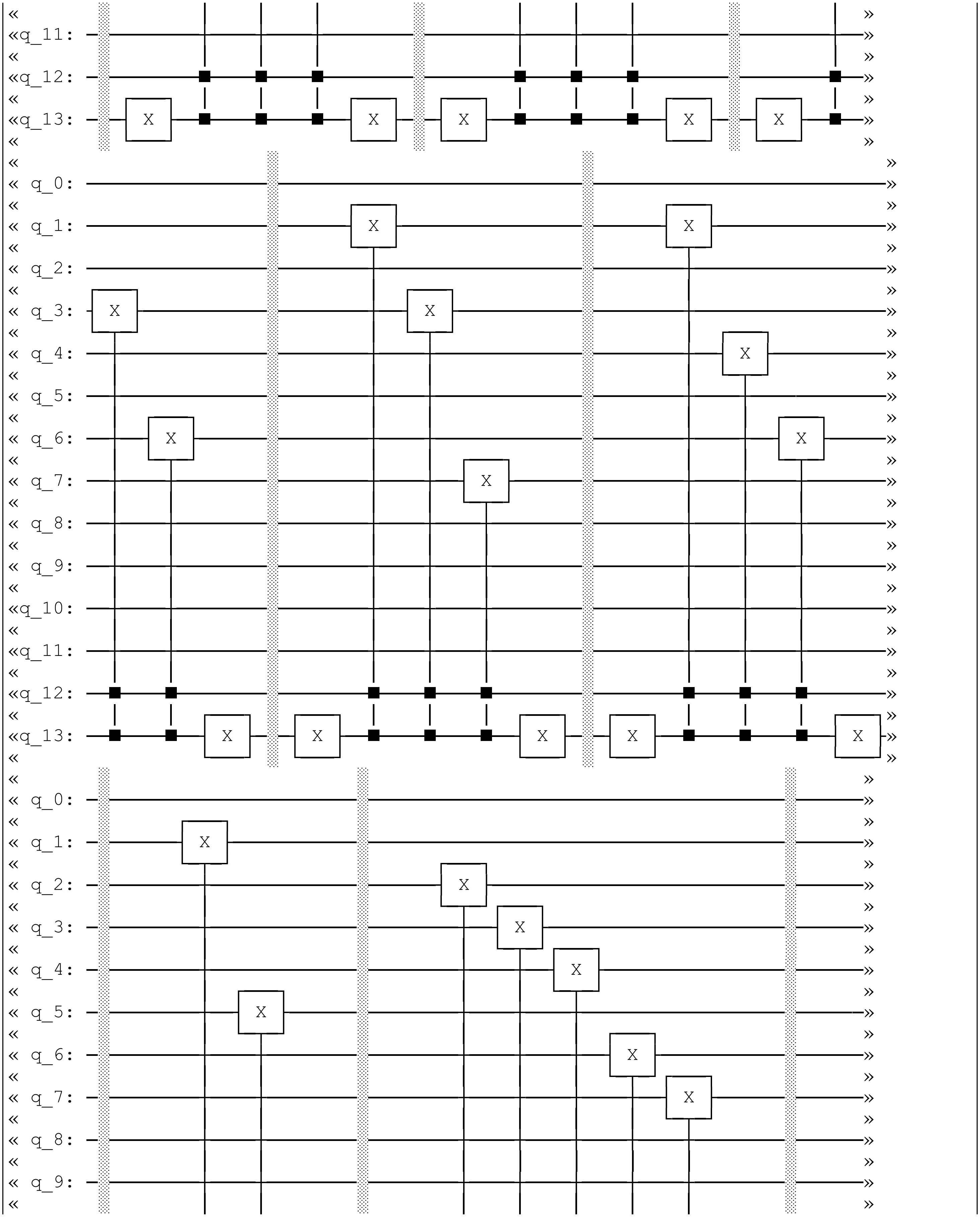

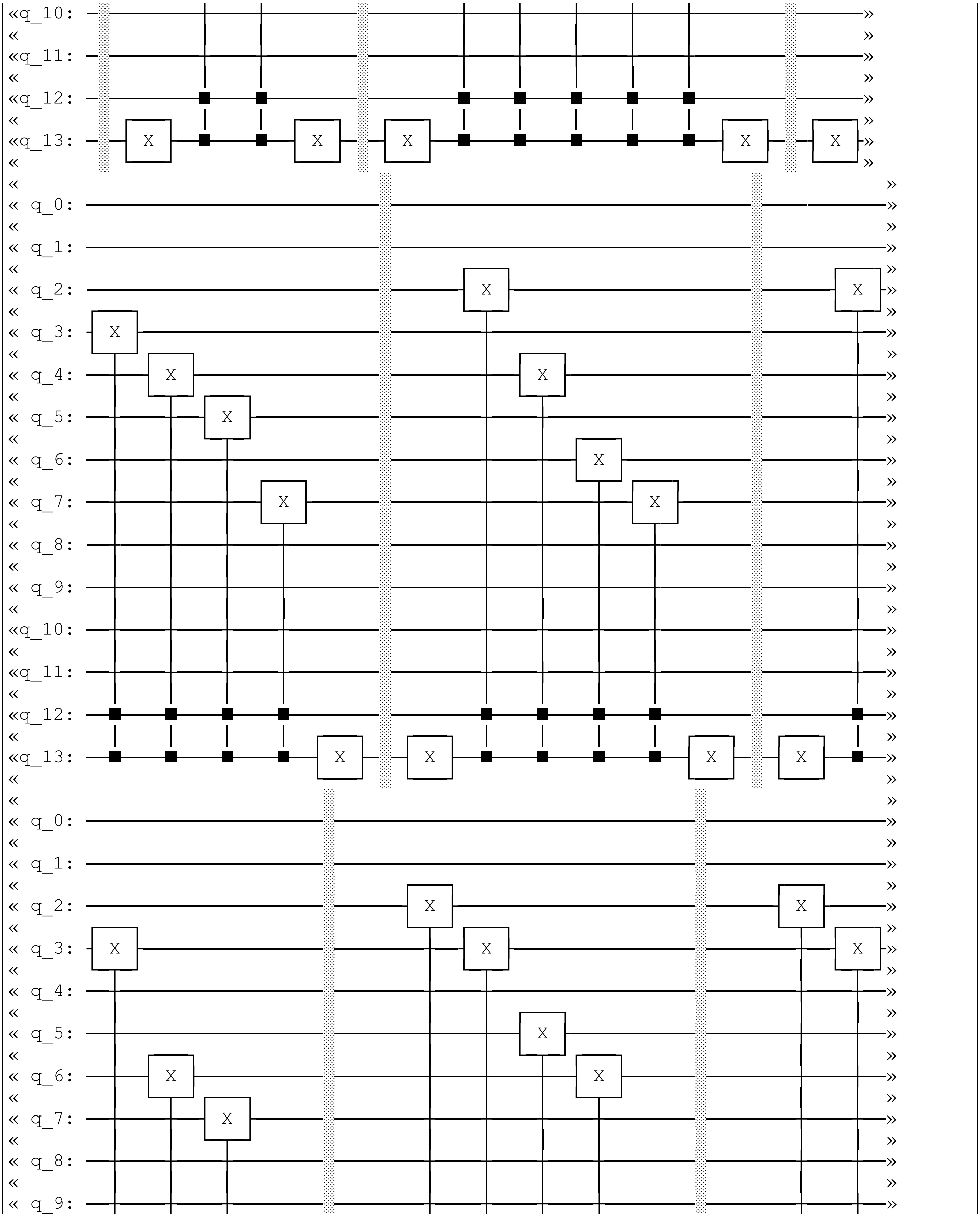

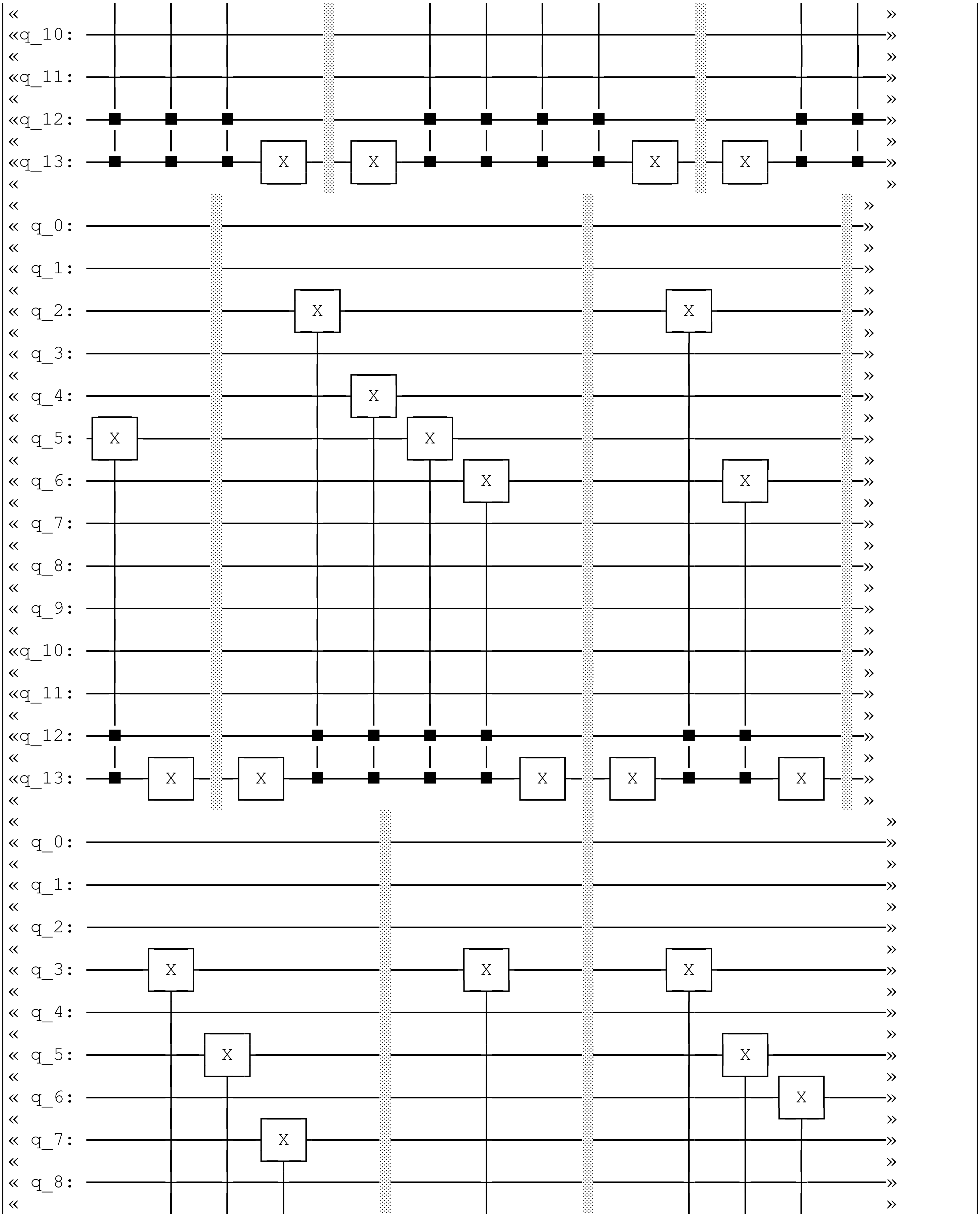

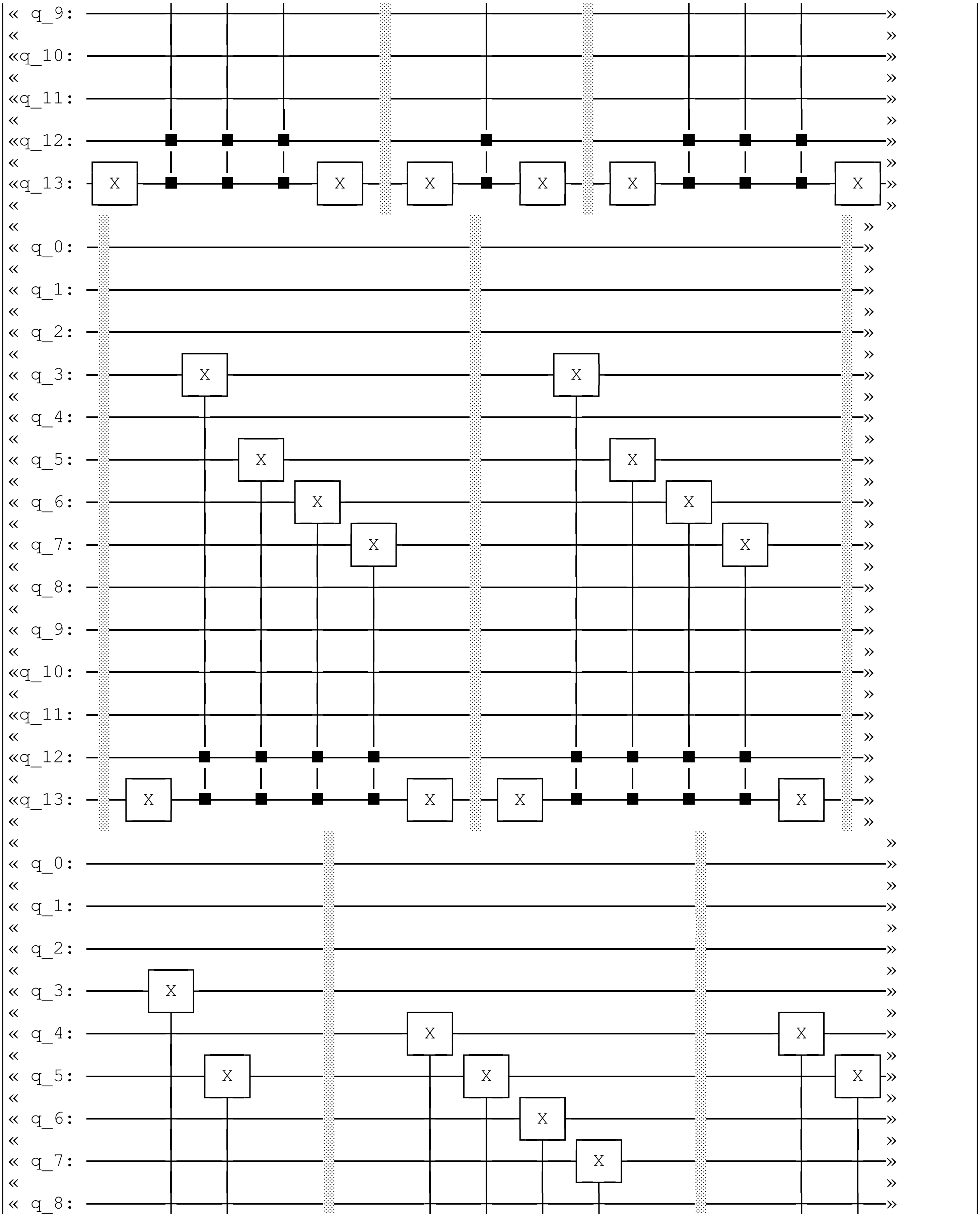

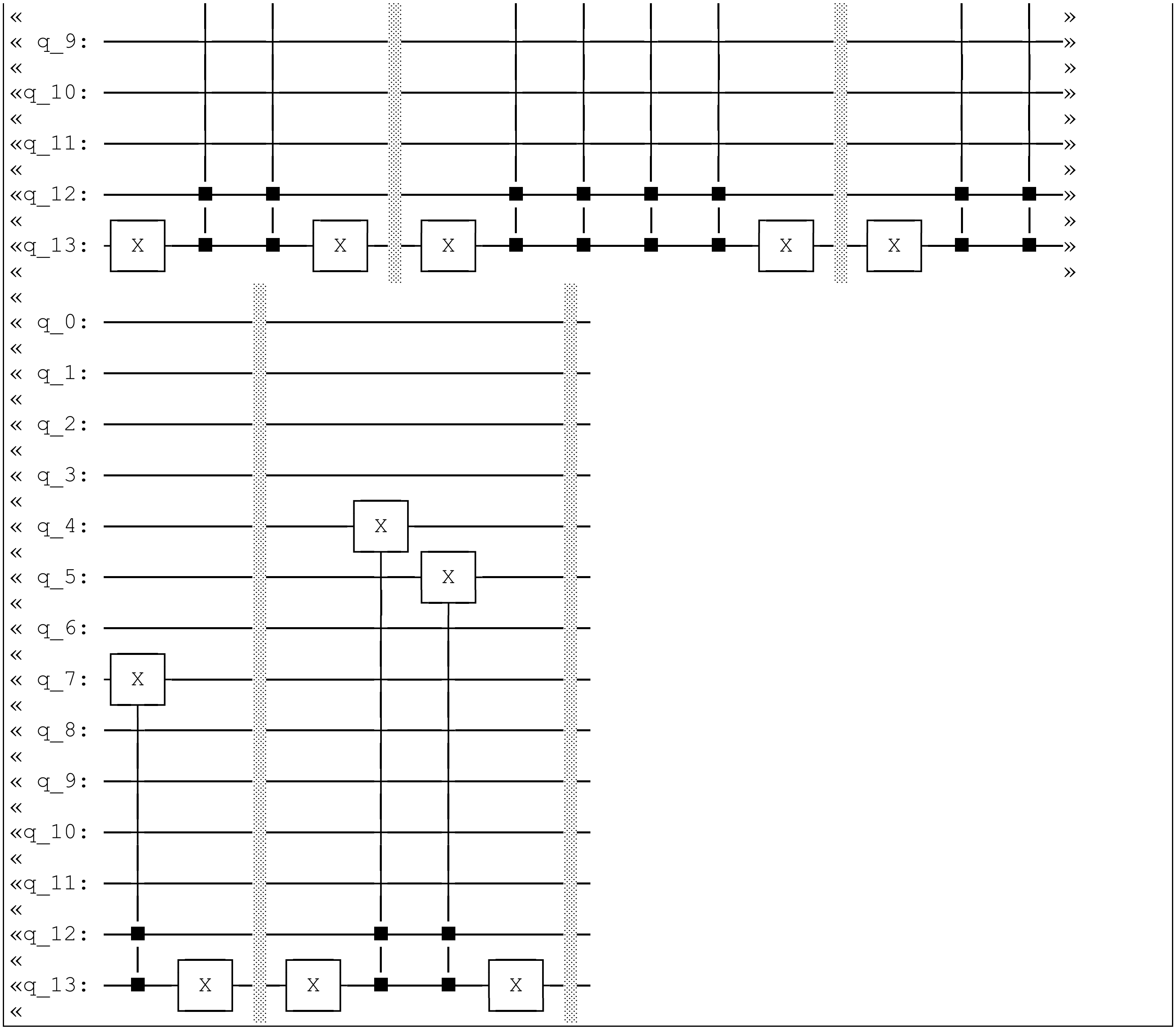

*Fig. A2. Circuit for the NEQR process*

## AUTHOR

First Author – Santanu Ganguly, *Cisco Systems International, UK; 9-11 New Square, Feltham TW14 8HA, United Kingdom*

ORCID iD: https://orcid.org/0000-0003-0141-9228

Correspondence Author – Santanu Ganguly, santagan@cisco.com , santanu.gangoly@gmail.com